\documentclass[twocolumn,twocolappendix]{aastex631}
\usepackage{amsmath,amssymb}
\makeatletter
\providecommand*{\approxident}{%
  \mathrel{%
    \mathpalette\@approxident\sim
  }%
}   
\newcommand*{\@approxident}[2]{%
  \sbox0{$#1\vcenter{}$}%
  \sbox2{$\m@th#1\equiv$}%
  \dimen2=\dimexpr\ht2 - \ht0\relax
  \sbox4{$\m@th#1\sim$}%
  \dimen4=\dimexpr\ht4 - \ht0\relax
  \dimen0=\dimexpr
    -\ht4 - \dp4 %
    + \dimen2 %
  \relax
  \vcenter{\offinterlineskip
    \copy4 %
    \kern\dimen0 %
    \copy4 %
    \kern\dimen0 %
    \copy4 %
    \ifdim\dp4=\z@
      \kern\dimexpr -\ht0 + \dimen4\relax
    \fi
  }%
}      
\makeatother

\shorttitle{Polarization with AIRCARS}
\shortauthors{Kansabanik et al.}

\begin{document}

\graphicspath{{./}}

\title{Tackling the Unique Challenges of Low-Frequency Solar-Polarimetry with the Square Kilometre Array Low Precursor : The Algorithm}

\author[0000-0001-8801-9635]{Devojyoti Kansabanik}
\affiliation{National Centre for Radio Astrophysics, Tata Institute of Fundamental Research, Pune University, Pune 411007, India}

\author[0000-0002-4768-9058]{Divya Oberoi}
\affiliation{National Centre for Radio Astrophysics, Tata Institute of Fundamental Research, Pune University Campus, Pune 411007, India}

\author[0000-0002-2325-5298]{Surajit Mondal}
\affiliation{Center for Solar-Terrestrial Research, New Jersey Institute of Technology, 323 M L King Jr Boulevard, Newark, NJ 07102-1982, USA}
\affiliation{National Centre for Radio Astrophysics, Tata Institute of Fundamental Research, Pune University, Pune 411007, India}

\correspondingauthor{Devojyoti Kansabanik}
\email{dkansabanik@ncra.tifr.res.in, devojyoti96@gmail.com}

\begin{abstract}
Coronal magnetic fields are well known to be one of the crucial parameters defining coronal physics and space weather. However, measuring the global coronal magnetic fields remains challenging. The polarization properties of coronal radio emissions are sensitive to coronal magnetic fields. While they can prove to be useful probes of coronal and heliospheric magnetic fields, their usage has been limited by technical and algorithmic challenges. We present a robust algorithm for precise polarization calibration and imaging of low-radio frequency solar observations and demonstrate it on data from the Murchison Widefield Array, a Square Kilometer Array (SKA) precursor. This algorithm is based on the {\it Measurement Equation} framework, which forms the basis of all modern radio interferometric calibration and imaging. It delivers high-dynamic-range and high-fidelity full-Stokes solar radio images with instrumental polarization leakages $<1\%$, on par with general astronomical radio imaging, and represents the state of the art. Opening up this rewarding, yet unexplored, phase space will enable multiple novel science investigations and offer considerable discovery potential. Examples include detection of low-level circularly polarization from thermal coronal emission to estimate large-scale quiescent coronal fields; polarization of faint gyrosynchrotron emissions from coronal mass ejections for robust estimation of plasma parameters; and detection of the first-ever linear polarization at these frequencies. This method has been developed with the SKA in mind and will enable a new era of high-fidelity spectropolarimetric snapshot solar imaging at low radio frequencies.
\end{abstract}

\accepted{10 April, 2022}\published{21 June, 2022}
\section{Introduction}\label{sec: introduction to the problem}
The solar corona is the outermost layer of the Sun and is made up of hot magnetized plasma. 
The magnetic field permeating this plasma couples the solar atmosphere to the solar interior.
Solar magnetic fields are responsible for the bulk of the observed coronal phenomenon, spanning a range of time scales from a solar cycle to flares lasting milliseconds and in terms of the energetics from massive coronal mass ejections (CMEs) to nanoflares.
Coronal magnetic fields also play a large role in shaping the coronal structures and even the solar wind.
Hence, to understand coronal dynamics it is essential to measure and understand the ever-evolving coronal magnetic fields.

The sun and the corona is routinely observed using multiple space-based and ground-based instruments, spanning from radio, visible, extreme ultraviolet (EUV), to X-ray wavelengths. 
Magnetic fields, however, are rather hard to measure \citep{Wiegelmann2014}. Routine measurements of global solar magnetic field are available only at the photosphere \citep{Lagg2017}. 
For lack of better options, large-scale quiescent coronal magnetic fields can only be estimated using extrapolations of photospheric magnetic fields \citep[e.g.][]{Wiegelmann2005,Jiang2014}.
During CME eruptions, EUV \citep[][etc.]{Gopalswamy_2011} and white light \citep[][etc.]{Schmidt2016} have also been used to measure the coronal magnetic fields. 
Very recently, the global coronal magnetic field  has been measured successfully using near-infrared observations in the range $1.05$--$1.35\ \mathrm{R_\odot}$, where $\mathrm{R_\odot}$ is the solar radius \citep{Yang2020}.
Measuring coronal magnetic fields is also a key science objective for the Daniel K. Inouye Solar Telescope \citep[DKIST;][]{Rast2021}, though these measurements are likely to remain limited to $<1.5\ \mathrm{R_\odot}$.

Under favorable circumstances, radio observations have also been used to estimate coronal magnetic fields associated with active regions and/or CMEs \citep{Alissandrakis2021}.
The coronal optical depth at these heights ($>1.3\ \mathrm{R_\odot}$) becomes too low for visible and EUV bands.
Although the radio observables, in principle, are sensitive to coronal magnetic fields, it has been technically too challenging to extract this information on a regular basis.
Most of the radio studies have focused on the active emissions, and the global quiescent coronal magnetic fields at high coronal heights have remained beyond reach.
    
The polarization properties of solar radio emission, in addition to being a direct probe of the coronal magnetic fields, can also provide strong constraints on the emission mechanisms. Despite its well-appreciated importance, low-frequency polarimetric observations of the Sun are one of the least explored area of solar physics.
The radio Sun is a complicated source. It has structures spanning a large range of angular scales. The spectral, temporal, and morphological characteristics of the radio emissions are also very dynamic. The brightness temperature ($\mathrm{T_B})$ of the low-frequency solar emissions can vary from $10^3\ \mathrm{K}$ for gyrosynchrotron emission from CME plasma \citep[e.g.][]{bastian2001,Mondal2020a} to $10^{13}\ \mathrm{K}$ for bright type III radio bursts \citep[e.g.][]{McLeanBook,Reid2014} over a background quiescent $\mathrm{T_B}$ of $\sim10^6\ \mathrm{K}$. Depending upon the emission mechanism at play, the polarization fraction can vary from $\leq 1\%$ to $\sim100\%$ \citep[e.g.][]{McLeanBook,Nindos2020}.

To date, most of the polarimetric studies of the Sun at low frequencies are based on nonimaging dynamic spectra measuring the circular polarization \citep[e.g.][]{Reid2014,Kaneda2017}. These observations cannot provide any information about the source structure or location. Some innovative instruments use simultaneous Stokes I imaging and Stokes V dynamic spectra to help in localization of the source of active emission \citep{Raja2014}. 
These studies implicitly assume the locations of the peaks in the total intensity and circularly polarized emission to be the same.
This assumption usually holds when there is a single dominant source of emission.
This approach is not useful when multiple sources of active emission are simultaneously present on the Sun \citep{Mohan2017spreads} or for weaker and/ or extended emission like gyrosynchrotron emission from CME plasma \citep[e.g.][]{bastian2001, Mondal2020a} and the free-free emission from the quiet Sun \citep{Sastry_2009}.

The variation in the solar emission over small temporal and spectral scales imposes a requirement for snapshot spectroscopic imaging. The need to be able to see features varying vastly in $\mathrm{T_B}$ highlights the need for a high imaging dynamic range. Only recently it has become possible to meet these exacting requirements for solar radio imaging. Several new generation low-frequency radio interferometers have now become available -- LOw Frequency ARray \citep[LOFAR;][]{lofar2013} operating at 10 -- 80 $\mathrm{MHz}$ and 120 -- 240 $\mathrm{MHz}$; Long Wavelength Array \citep[LWA;][]{kassim2010} at 10 -- 88 $\mathrm{MHz}$; and the Murchison Widefield Array \citep[MWA;][]{Tingay2013,Wayth2018} at 80 -- 300 $\mathrm{MHz}$. Of these, the MWA has a large number of antenna elements (a total of 128 antenna elements) spread over a small footprint in a centrally condensed configuration and is especially well suited for high-dynamic-range high-fidelity snapshot spectroscopic imaging. Even with the availability of data from a capable instrument like the MWA, there remains another problem to overcome -- that of being able to produce the large number of high-quality images necessary to enable the science in a reliable unsupervised manner. This need was met by a robust Stokes I calibration and imaging pipeline developed by \cite{Mondal2019} designed to build on the strengths of the MWA architecture. This pipeline is christened Automated Imaging Routine for Compact Arrays of the Radio Sun (AIRCARS) and the high-fidelity spectroscopic snapshot images it delivers now represent the state of the art and have already led to new discoveries such as quasiperiodic pulsations of solar radio bursts \citep{Mohan2019a,Mohan2019b,Mondal2021a}, weak nonthermal emissions from the quiet Sun \citep{Mondal2020b,Mondal2021b,Mohan2021a,Mohan2021b}, and weak gyrosynchrotron emission from CMEs \citep{Mondal2020a}. 
Having established the ability of MWA solar observations to deliver high-fidelity Stokes I images, it is now the right time to use that experience for high-fidelity polarization imaging.

Full-Stokes calibration is significantly more challenging than working with Stokes I alone.
These challenges are even greater for the case of low radio frequency solar imaging.
On the one hand, solar emission can have a very large range of intrinsic polarizations, which can also vary rapidly across time and frequency. The MWA has a large field of view (FoV). Based on the FWHM of the primary beam, at 150 {$\mathrm{MHz}$} the FoV of the MWA is $\sim$610 $\mathrm{degree^2}$, which reduces to $\sim$375 $\mathrm{degree^2}$ by 200 $\mathrm{MHz}$ \citep{Tingay2013}. The wide FoV aperture arrays tend to have large instrumental polarization imposing a strong requirement for precise calibration.
In fact, some of the assumptions made for routine polarimetric calibration at higher frequencies for small FoV instruments no longer hold in this regime \citep{lenc2017}.
We present a general algorithm for polarimetric calibration suitable for our application.
We have implemented this algorithm and demonstrate its efficacy on the MWA solar data.
The algorithm will be well suited for solar imaging with the future SKA-Low and other interferometers with centrally condensed array configurations.

The paper is organized as follows.
We first briefly discuss some basics of polarization calibration in Section \ref{sec : basic_polarimetry} to build up the base for our calibration algorithm. Section \ref{sec : challenges} describes the challenges of the polarization calibration of the Sun at low frequencies and the limitations of the conventional methods of polarization calibration. We then describe the new algorithm in Section \ref{Overview of the Algorithm}. Section \ref{sec : result} presents the final results with a discussion, and Section \ref{Conclusion} provides the conclusion.

\section{polarization calibration of a radio interferometer}\label{sec : basic_polarimetry}
A radio interferometer is made up of a number of radio antennas or elements.
These antennas measure the voltages corresponding to the two orthogonal polarizations of the electric field, $\Vec{\mathrm{E}}$, incident on the antenna.
$\Vec{\mathrm{E}}$ could be measured in either of the linear or circular bases -- ($\mathrm{E_X},\ \mathrm{E_Y}$) or ($\mathrm{E_R},\ \mathrm{E_L}$) respectively. Incident $\Vec{\mathrm{E}}$ induces a voltage in the antenna, and the primary observable of an interferometer is the cross correlation between the components of the induced voltages for every antenna pair, referred to as {\it visibilities}. To capture the complete information about the state of polarization of $\Vec{\mathrm{E}}$, for any given antenna pair described by indices $\mathrm{i}$ and $\mathrm{j}$, an interferometer needs to measure  $\mathrm{X_iX_j^\dagger},\ \mathrm{X_iY_j^\dagger},\ \mathrm{Y_iX_j^\dagger},\ \mathrm{Y_iY_j^\dagger}$ (or equivalently $\mathrm{R_iR_j^\dagger},\ \mathrm{R_iL_j^\dagger},\ \mathrm{L_iR_j^\dagger},\ \mathrm{L_iL_j^\dagger}$) {\footnote{$\dagger$ represents conjugate transpose.}}. This complete set of visibilities is often referred to as {\it full-polar} visibilities.

The measured visibilities include the corruption due to atmospheric propagation effects and instrumental effects. 
In order to arrive at the true visibilities corresponding to the astronomical sources, these corruptions need to be removed.
This process is known as calibration.
\cite{Hamaker1996_1} proposed a general framework for polarimetric calibration based on the {\it measurement equation}.
Briefly, the measured complex voltage vector, $\Vec{\mathrm{V}}$, per antenna can be expressed in terms of $\Vec{\mathrm{E}}$, and the antenna-based Jones matrix, $\mathrm{J}$, \citep{Jones1941} as:
\begin{equation}\label{eq:jones_matrix}
\begin{split}
    \Vec{\mathrm{V}} &=\mathrm{J}\ \Vec{\mathrm{E}}\\
    \begin{pmatrix}\mathrm{X}\\\mathrm{Y}\end{pmatrix} &=\mathrm{J} \begin{pmatrix}\mathrm{E_X}\\\mathrm{E_Y}\end{pmatrix},
\end{split}
\end{equation}
where $\mathrm{J}$ is a $2\times2$ matrix representing the instrumental and atmospheric effects. 
The measured correlation products for two antennas represented by indices $\mathrm{i}$ and $\mathrm{j}$ can be written as a $2\times2$ matrix, also known as the \newline {\it Visibility matrix}, as follows \citep{Hamaker2000,Smirnov2011}:
\begin{equation}\label{eq:visibility_matrix}
    \begin{split}
      { \mathrm{V_{ij}}^\prime}&= 2\begin{pmatrix} \mathrm{X_iX_j^\dagger} & \mathrm{X_iY_j^\dagger}\\\mathrm{Y_iX_j^\dagger} & \mathrm{Y_iY_j^\dagger}
        \end{pmatrix}\\
       \mathrm{V_{ij}^\prime}&=2\ \mathrm{J_i}
        \begin{pmatrix}
        \mathrm{E_{X,i}E_{X,j}}^\dagger & \mathrm{E_{X,i}E_{Y,j}^\dagger}\\\mathrm{E_{Y,i}E_{X,j}^\dagger} & \mathrm{E_{Y,i}E_{Y,j}^\dagger}
        \end{pmatrix}
       \mathrm{J_j^\dagger}\\
       \mathrm{V_{ij}^\prime}&=\mathrm{J_i}\begin{pmatrix}
       \mathrm{V_I+V_Q} & \mathrm{V_U+iV_V} \\ \mathrm{V_U-iV_V} & \mathrm{V_I-V_Q}
       \end{pmatrix}_\mathrm{ij}\mathrm{J_j^\dagger}\\
      \mathrm{V_{ij}^\prime}&= \mathrm{J_i\ V_{ij}\ J_j^\dagger},
    \end{split}
\end{equation}
where $\mathrm{V_{ij}}$ is the true source visibility matrix; $\mathrm{V_{ij}^\prime}$ is the observed visibility matrix; and  $\mathrm{V_I},\ \mathrm{V_Q},\ \mathrm{V_U}$ and $\mathrm{V_V}$ are the Stokes visibilities of the incident radiation. 
Here we have followed the IAU/IEEE definition of the Stokes parameters \citep{IAU_1973,Hamaker1996_3}. The four Stokes parameters, $\mathrm{I,\ Q,\ U,\ V}$ were originally defined by \citet{stokes1851}. Stokes I represents the total intensity; Stokes Q and U represents the linear polarization, and circular polarization is denoted by Stokes V. All $\mathrm{V_{ij}^\prime}$s have independent additive noise, $\mathrm{N_{ij}}$, associated with them, and the Eq. \ref{eq:visibility_matrix} can be written as:
\begin{equation}\label{eq:measurement_equation}
    \begin{split}
         \mathrm{V_{ij}^\prime}&=\mathrm{J_i\ V_{ij}\ J_j^\dagger + N_{ij}}.
    \end{split}
\end{equation}
Equation \ref{eq:measurement_equation} is referred to in literature as the {\it measurement equation} of a radio interferometer \citep{Hamaker1996_1,Hamaker2000,Smirnov2011}.

The objective of polarization calibration is to estimate $\mathrm{J_i}$s for all antennas and obtain $\mathrm{V_{ij}}$ from the $\mathrm{V_{ij}^\prime}$. 
This requires correcting for four different aspects.
These aspects and their impacts are enumerated below.
\begin{enumerate}
\item Time-variable instrumental gain: While they are independent in origin, in practice, it is not feasible to disentangle the atmospheric propagation effects from time-variable instrument gains. So these effects are clubbed with instrumental gains in this formalism. 
The impact of these gain variations is to make the interferometer incoherent.
\item Frequency-dependent instrumental bandpass: This can modify the true spectral signature of the source and introduce incoherence across the frequency axis.
\item Polconversion: This can lead to a leakage from Stokes I to other Stokes parameters and thus modifies the observed magnitude of the observed polarization vector, $\mathrm{p=(Q,\ U,\ V)}$.
\item Polrotation: This is the mixing between Stokes Q, U, and V and leads to a rotation of $\mathrm{p}$.
\end{enumerate}

The time and frequency dependence of the instrumental gains arise due to the nature of the signal chain and the atmospheric propagation effects. 
These are routinely corrected for in standard interferometric calibration.
Ideally a set of orthogonal receptors are expected to receive only the matched orthogonal component of the incident $\vec{E}$.
In practice, reasons ranging from proximity to other receptors, manufacturing tolerances to cross talk between closely placed cables, imply that a given orthogonal receptor also picks up some amount of signal of the other component of $\vec{E}$.
This mixing of the orthogonal components of $\vec{E}$ gives rise to {\it polconversion}.
Things like misalignment of the dipoles with respect to the sky coordinates and the phase differences between the two orthogonal receptors give rise to {\it polrotation}. 

\section{Challenges and limitations}
\label{sec : challenges}
A general mathematical foundation of polarization calibration has been provided by \citet{Hamaker2000} and a more recent review is available in \citet{Smirnov2011}.
While a general prescription has been available, the complexity of the problem and its computation-heavy nature have restricted most commonly available implementations to make some simplifying assumptions.
This is an active area of research, especially in view of the upcoming ambitious facilities like the SKA, and new algorithms and implementations are being developed by multiple groups across the world \citep[e.g.][etc.]{Smirnov2011,Mitchell2008,stefcal2014,cubical2018}.
This section lists the challenges of low-frequency solar observation and the limitation of conventional algorithms and earlier attempts.

\subsection{Challenges of Low-frequency Solar Observations}
These challenges are related to the large FoVs and the nature of low-frequency aperture array instruments. Conventional methods correct for polconversion using observations of a strong unpolarized calibrator source. Correcting for polrotation requires observations of a strong source with known polarization properties \citep{Hamaker2000,Hales2017}.

The large FoV of low radio frequency aperture array instruments imply that:
\begin{enumerate}
    \item As there are no moving parts and the beam is steered electronically, the primary beam of instrument can vary dramatically with pointing direction.
    \item Given the large FoV and nature of low radio frequency sky, there is no single polarized source strong enough to dominate the observed $\mathrm{V_{ij}^{\prime}}$.
\end{enumerate}
This imposes the requirement that, for good polconversion calibration, the target field and the calibrator should be observed with the same pointing, which is rarely the case.
The lack of a dominant polarized source makes it hard to do polrotation calibration.
In addition, the low radio frequencies of observation require one to contend with the direction-dependent ionospheric distortion and Faraday rotation (FR).
A comprehensive discussion of a successful approach to deal with these challenges has been provided by \citet{lenc2017}.

In addition to the ones stated above, solar observations at these frequencies have additional challenges to deal with.
These include:
\begin{enumerate}
    \item The large flux density of the Sun and the wide FoVs imply that daytime calibrator observations are corrupted by solar contributions. 
    Hence calibrators for solar observations are usually observed before sunrise or after sunset. 
    The large time gap between the calibrator and the source observations, in addition to the difference in the pointing direction, reduce our ability to constrain the true state of the instrument and the ionosphere.
    This increases our reliance on self-calibration-based methods to obtain high-dynamic-range solar radio images.
    \item Solar emission is not assured to be unpolarized and can vary significantly across time and frequency.
    In the conventional approach, this makes the Sun unsuitable as a calibrator source for polconversion and polrotation.
\end{enumerate}

\subsection{Limitations of Conventional Algorithms}
\label{sec:conventional-algos}
Most of the standard interferometric calibration and imaging packages like {\sc CASA} and {\sc AIPS} implement a linearized {\it measurement equation}. While they have been spectacularly successful in delivering high-quality polarimetric image, the following two assumptions must be satisfied for this formalism to be valid:
\begin{enumerate}
    \item The instrumental polarization must be small \newline($< 10$\%).
    \item The fractional polarization of the sources used for calibration should be low \newline($< 10$\%).
\end{enumerate}
In the usual case of steerable antennas where the FoVs are small enough that one is never too far from the optical axis of the dish, the instrumental polarization usually meets this threshold.
Also, the fractional polarization of standard polarization calibrators (e.g. 3C 286, 3C 138, 3C 48) is $\leq10\%$. A linearized formalism is, hence, quite adequate for most applications. For our particular application of looking at the Sun with an aperture array, however, neither of these assumptions hold true. 
The instrumental polarization is a strong function of the pointing direction. It increases as one goes farther from the zenith and/or cardinal directions and is often much larger than 10\%. The solar emission can vary dramatically in its intrinsic polarization from being unpolarized to being nearly 100\% polarized. 
As calibrator observations are not possible during the solar observation with the MWA, one has to rely on the self-calibration using the Sun itself. The potentially very large fractional polarization of solar emission implies that one cannot rely on using them for polarization self-calibration with the linearized algorithms.
This forces us to develop a more general formalism for polarimetric calibration.

The linear approximation also restricts the dynamic range of the Stokes images \citep{Smirnov2011}. In addition, in the linearized formulation, the instrumental leakages are calibrated only using the cross-hand visibilities ($\mathrm{XY^\dagger,\ YX^\dagger}$ or $\mathrm{RL^\dagger,\ LR^\dagger}$) \citep{Hales2017}, and the parallel-hand visibilities ($\mathrm{XX^\dagger,\ YY^\dagger}$ or $\mathrm{RR^\dagger,\ LL^\dagger}$) are simply ignored. Hence, while the cross-hand visibilities are updated during polarization self-calibration, the parallel-hand correlations remain unchanged. As a consequence this algorithm is unsuitable for iterative implementation.

\subsection{Previous Attempts and Their Limitations}
\label{sec:previous-attempts}
While \citet{Mondal2019} have used self-calibration-based methods with remarkable success to obtain high-dynamic-range solar Stokes I images, their algorithm does not include polarimetric imaging. It also does not include absolute flux density calibration, which needs to be done independently \citep{Kansabanik2022}.
\citet{Patrick2019} have demonstrated polarimetric solar radio imaging with the MWA. They used nighttime calibrator observations to estimate the instrumental and ionospheric gains.
They used an ad-hoc approach to mitigate instrumental polarization, which we refer to as Method I in the following text.
The assumptions and requirements of Method I, and the limitations they impose are listed below:
\begin{enumerate}
    \item The leakage from Stokes I to other Stokes components remains essentially constant across the angular span of the solar disk. While this assumption is valid for some pointing, it is not true in general and this is demonstrated in detail in Sec. \ref{sec:ideal_beam}.
    \item $\mathrm{S_Q=\ S_U=\ 0}$, where $\mathrm{S_Q}$ and $\mathrm{S_U}$ represent the Stokes Q and U components of the solar flux density, i.e. the linearly polarized emission from the Sun is assumed to be exactly zero.
    While no linearly polarized emission has been reported from the Sun at low radio frequencies yet, the new generation of instruments can now provide spectroscopic snapshot images over spectral spans as small a few $\mathrm{kHz}$, as opposed to order $\mathrm{MHz}$ available earlier. Assuming the linearly polarized flux density to be zero precludes their discovery and locks us out of an interesting discovery phase space.
    \item Method I relies on the fact that the fractional circular polarization from the quiet Sun is expected to be small. 
    It attempts to estimate an epoch-dependent instrumental leakage by minimizing the total number of pixels that show a fractional circular polarization larger than a chosen threshold, $\mathrm{r_c}$. It inherently assumes that this emission is coming from quiet-Sun regions.
    At low radio frequencies, the presence of multiple simultaneous active sources on the Sun can, however,
    limit the regions of the solar disk with quiet Sun emission \citep{Mohan2017spreads}. Even though the area occupied by the active regions is a small fraction of the solar disk at optical and higher frequencies, at low radio frequencies even the smallest active region gets scatter broadened to a few arcmin \citep[e.g.,][]{Kontar2017,Mohan2021b}.
    In addition, the intrinsic brightness temperatures, $\mathrm{T_B}$, of various emissions often found to appear simultaneously on the Sun vary from $10^4\ \mathrm{K}$ for gyrosynchrotron emission to up to  $10^{13}\ \mathrm{K}$ for type III solar radio bursts. The presence of very bright nonthermal emission has two consequences -- one they lead to an increase in the system temperature and consequently the thermal noise in the image; and two they impose a larger imaging dynamic range requirement to be able to image the $\lesssim 10^6\ \mathrm{K}$ quiet-Sun regions in the presence of much brighter nonthermal emission.
\end{enumerate}

The primary merit of Method I is that it enabled the authors to get to interesting science using an approximate and quick correction of instrumental polarization and circumventing the effort and complexities of developing and implementing a formally correct polarimetric calibration algorithm \citep{Patrick2019, Rahman2020}. The ionospheric phases during the solar observations are expected to be significantly different compared to those determined from nighttime calibrator observation. Hence applying the gain solutions from the nighttime calibrator observations limits the images to much poorer fidelity than the intrinsic capability of the data. These images cannot be used for reliable measurements of low levels of circular polarization. Method I is also known to give rise to some spurious polarization for very bright solar radio bursts \citep{Rahman2020}.

\section{Algorithm}\label{Overview of the Algorithm}
This section describes a robust formal polarization calibration algorithm that overcomes the shortcomings mentioned in Sec. \ref{sec:previous-attempts} and enables high-fidelity polarimetric imaging.
It builds on three pillars; i) self-calibration, ii) availability of a reliable instrumental beam model, and iii) some well-established properties of low-frequency solar radio emission. For low-radio-frequency solar observations with aperture arrays, it is not feasible to obtain calibrator observations at nearby times with the same primary beam pointing as used for target observations. This algorithm is, hence, designed to not require any calibrator observations. We refer to this algorithm as Polarimetry using Automated Imaging Routine for Compact Arrays for the Radio Sun \newline (P-AIRCARS). 
A detailed description of the algorithm is presented here, and that of its implementation as a robust unsupervised pipeline will be the subject of a forthcoming paper (D. Kansabanik et al. 2022, in preparation).

\subsection{Full Jones calibration algorithm}
In conventional polarization calibration tools like {\sc CASA}, all four observed visibilities between an antenna pair are written as separate equations in terms of instrumental gains and leakages. These equations are approximated up to first-order terms in leakages and solved separately to obtain the instrumental parameters \citep[e.g.][]{Hales2017}. In full Jones calibration, the {\it Measurement Equation} is solved as a $2\times2$ matrix equation. From the Eq. \ref{eq:visibility_matrix} we can write the coherence noise, $\mathrm{S}$, as,
\begin{equation}\label{eq:coherency_noise}
\begin{split}
     \mathrm{S}&=\mathrm{\sum_{ij} ||J_i^{-1}V_{ij}^\prime J_j^{\dagger{-1}} - V_{ij}||^2_F}\\
     \mathrm{S}&=\mathrm{\sum_{ij} Tr\left[ \left( J_i^{-1}V_{ij}^\prime J_i^{\dagger-1}-V_{ij}\right) \left(J_i^{-1}V_{ij}^\prime J_i^{\dagger-1}-V_{ij}\right)^\dagger \right]}
\end{split}
\end{equation}
where $\mathrm{||.||_F}$ represents the Frobenius norm \citep{horn_johnson_1985} of the matrix, $\mathrm{V_{ij}}$ is the model visibility, $\mathrm{V_{ij}}^\prime$ the observed visibility, and $\mathrm{J_i}$ and $\mathrm{J_j}$ represent Jones matrices for antennas $\mathrm{i}$ and $\mathrm{j}$ respectively. The instrumental Jones matrices are estimated by minimizing $\mathrm{S}$. Minimization of $\mathrm{S}$ leads to the matrix generalization of the conventional scalar calibration, which has been used in several standard interferometric software packages like CASA \citep{mcmullin2007}, AIPS \citep{Wells1985}, flagcal \citep{Flagcal2012}, and classical {\it antsol} 
\citep{Bhatnagar2001}, where Jones matrices were replaced by a single complex number.
The full Jones calibration was introduced by \citet{Hamaker2000} and \citet{Mitchell2008} and was later optimized as {\it StefCal} \citep{stefcal2014}. We have used the recently developed full Jones calibration software package, {\it CubiCal} \citep{cubical2018,Cubical_robust2019}, which uses complex optimization and the Wirtinger derivative \citep{Wirtinger1927}. In brief, minimization of $\mathrm{S}$ reduces to an analytical update rule of $\mathrm{J_i}$ in terms of $\mathrm{V_{ij},V_{ij}^\prime}$ and the Jones matrices of other antennas as,
\begin{equation}\label{eq:jones_update}
    \mathrm{J_i^{\dagger{-1}}}=\mathrm{\left[ \sum_j V_{ij}J_j^{-1}V_{ij}^{\prime\dagger}\right] \left[ \sum_j V_{ij}J_j^{\dagger{-1}}J_j^{-1}V_{ij}^\dagger\right] ^{-1}}
\end{equation}
The Jones matrices of all the antenna elements are initialized as the identity matrix. $\mathrm{J_i}$s are estimated using Equation \ref{eq:jones_update} in subsequent iterations until the absolute value of the changes in the Jones terms and $\mathrm{S}$ fall below a small positive number ($\epsilon \sim 10^{-6}$) between two consecutive iterations. We find that solutions generally converge within $\sim$20 -- 30 iterations. 

\subsection{Self-calibration algorithm of P-AIRCARS}\label{paircars_algorithm}
In radio interferometry, it is standard practice to write the instrumental Jones matrices as a chain of independent 2$\times$2 matrices, each with its distinct physical origin and referred to as the Jones chain \citep{Smirnov2011}. In P-AIRCARS, the net Jones matrix for the $i$th antenna is given by,
\begin{equation}\label{eq:jones_terms}
\begin{split}
\mathrm{J_i(\nu,\ t,\ \Vec{l}\ )}&=\mathrm{G_i(t,\ \vec{l}
)\ B_i(\nu)\ K_{\mathrm{cross}}(\nu,\ t)\ D_i(\nu,\ t,\ \vec{l})}\\
&\times \mathrm{E_i(\nu,\ t,\ \Vec{l}\ )}
\end{split}
\end{equation}
where $\nu$, $t$ and $\vec{l}$ refer to the frequency and time of observation and direction ($\theta, \phi$) respectively.

These individual terms in Eq. \ref{eq:jones_terms} for antenna $i$ are:
\begin{enumerate}
\item $\mathrm{G_i(t,\ \vec{l})} = \begin{pmatrix}
\mathrm{g_{i,X}(t,\ \vec{l})} & 0\\ 0 & \mathrm{g_{i,Y}(t,\ \vec{l})}
\end{pmatrix}$ represents the frequency-independent time-variable instrumental gain.
This term also includes the direction-dependent gains arising due to propagation effects. 
At low radio frequencies, they correspond primarily to ionospheric phase.
\item $\mathrm{B_i(\nu)} = \begin{pmatrix}
\mathrm{b_{i,X}(\nu)} & 0\\ 0 & \mathrm{b_{i,Y}(\nu)}
\end{pmatrix}$ represents the time-independent instrumental bandpass response.
\item $\mathrm{K_{cross}}(\nu,\ t) = \begin{pmatrix}
e^{i\frac{\psi(\nu,\ t)}{2}} & 0\\ 0 & e^{-i\frac{\psi(\nu,\ t)}{2}}
\end{pmatrix}$ is the Jones matrix representing the phase difference between the two orthogonal receptors for the reference antenna,  given by $\psi(\nu, t)$.
\item $\mathrm{E_i}(\nu,\ t,\ \Vec{l}) = \mathrm{E_i}(\nu,\ t,\ \theta,\ \phi) \\= \begin{pmatrix}
\mathrm{E_{i,X\theta}(\nu,\ t, \theta)} & \mathrm{E_{i,X\phi}(\nu,\ t,\ \phi)}\\\mathrm{E_{i,Y\theta}(\nu,\ t, \theta)} & \mathrm{E_{i,Y\phi}(\nu,\ t,\ \phi)}
\end{pmatrix}$ is the direction-dependent modeled instrumental primary beam response.
\item $\mathrm{D_i}(\nu,\ t,\ \vec{l}) = \begin{pmatrix}
\mathrm{d_{i,XX}}(\nu,\ t,\ \vec{l}) & \mathrm{d_{i,XY}}(\nu,\ t,\ \vec{l})\\ \mathrm{d_{i,YX}}(\nu,\ t,\ \vec{l}) & \mathrm{d_{i,YY}}(\nu,\ t,\ \vec{l})
\end{pmatrix}$ is the error on the ideal instrumental primary beam model.
\end{enumerate}
The Sun being by far the dominant source in the sky implies that we are in essentially a small FoV regime.
This offers the advantage that the direction dependence due to the ionospheric effects included on $\mathrm{G_i(t,}\, \ \vec{l})$ can be ignored. We have verified that the direction dependence of $\mathrm{E_i(\nu,\ t,}\ \vec{l})$ cannot be ignored over the angular span of the Sun, but that of the $\mathrm{D_i(\nu,\ t,}\ \Vec{l})$ can be, as discussed in detail in Sec. \ref{subsec:imaged_based_cor}. 
Henceforth in this work, we ignore their direction dependence and regard $\mathrm{G_i(t,}\, \ \vec{l}) \approxident \mathrm{G_i(t)}$ and $\mathrm{D_i(\nu,\ t,}\ \Vec{l}) \approxident \mathrm{D_i(\nu,\ t)}$.

\begin{figure*}[!t]
    \centering
    \includegraphics[trim={0.8cm 1.5cm 0.0cm 1cm},clip,scale=0.62]{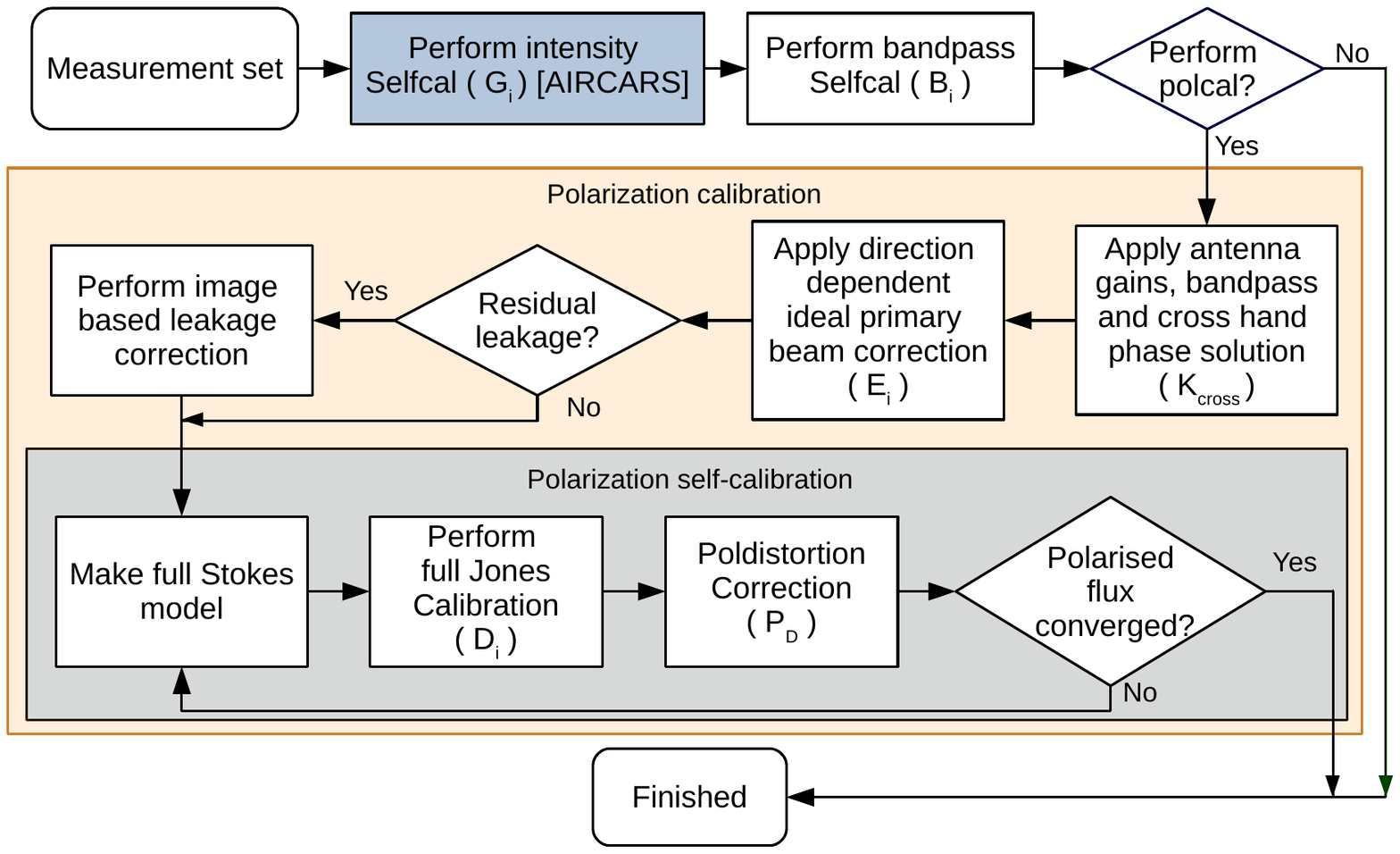}
    \caption{{\textbf{Flowchart describing the self-calibration algorithm of P-AIRCARS.}} The first major calibration block is intensity self-calibration, which is similar to that implemented in AIRCARS (marked in blue). The orange shaded box represents the polarization calibration block. The gray shaded box is a subset of the polarization calibration and shows the steps for polarization self-calibration. The Jones terms in Eq. \ref{eq:jones_terms} is being solved at each steps are denoted inside brackets.}
    \label{fig:polcal_flowchart}
\end{figure*}

If we write the $\mathrm{V_{ij}}$ in terms the of sky brightness matrix, $\mathbf{B}(\Vec{l})$ using the van Cittert-Zernike theorem \citep{thompson2017}, and neglecting the noise term, Eq. \ref{eq:measurement_equation} can be written as ,
\begin{equation}\label{eq:vcz}
\begin{split}
 \mathrm{V_{ij}^\prime}=&\mathrm{J_i}\ \left[ \iint \mathbf{B}(\nu,\ t,\ \Vec{l})\right.\\
 &\left.\times \mathrm{e^{-2\pi i(u_{ij}l+v_{ij}m+w_{ij}(n-1))}}\frac{\mathrm{dl\ dm}}{\mathrm{n}} \right]\ \mathrm{J_j^\dagger}
\end{split}
\end{equation}
where $\mathrm{l,\ m}$ and $\mathrm{n}$ are the direction cosines of $\Vec{l}$; and $\mathrm{u,\ v}$, and $\mathrm{w}$ are the components of the baseline vector in units of the wavelength. Using Eqs. \ref{eq:jones_terms} and \ref{eq:vcz}, we can write $\mathrm{V_{ij}}$ as,
\begin{equation}\label{eq:decomposed_me}
\begin{split}
\mathrm{V_{ij}^\prime\ (\nu,\ t)}=&\ \mathrm{G_i(t)\ B_{i}(\nu)\ K_{\mathrm{cross}}(\nu,\ t)\ D_{i}(\nu,\ t)}\\
 &\times \left[ \iint \mathrm{E_{i}(\nu,\ t,}\ \vec{l})\ \mathbf{B}(\nu,\ t,\ \Vec{l})\ \mathrm{E_{j}^\dagger(\nu,\ t,}\ \Vec{l})\right. \\
 &\times \left. \mathrm{e^{-2\pi i(u_{ij}l+v_{ij}m+w_{ij}(n-1))}}\frac{\mathrm{dl\ dm}}{\mathrm{n}} \right] \\
 &\times \mathrm{D_j^\dagger(\nu,\ t)\ K_{\mathrm{cross}}^\dagger(\nu,\ t)\ B_{j}^\dagger(\nu)\ G_{j}^\dagger(t)}\\
\mathrm{V_{ij}^\prime\ (\nu,\ t)}=&\ \mathrm{G_i(t)\ B_{i}(\nu)\ K_{\mathrm{cross}}(\nu,\ t)\ D_{i}(\nu,\ t)\ V_{ij,app}(\nu,\ t)}\\
&\times \mathrm{D_j^\dagger(\nu,\ t)\ K_{\mathrm{cross}}^\dagger(\nu,\ t)\ B_{j}^\dagger(\nu)\ G_{j}^\dagger(t)}
\end{split}
\end{equation}
Here the quantities outside the square bracket are the direction-independent Jones terms and the quantities inside the square brackets are the direction-dependent terms. We estimate each of these Jones terms step-by-step. The flowchart of the P-AIRCARS algorithm for estimating these Jones terms is shown in Fig. \ref{fig:polcal_flowchart}.

\subsection{Intensity self-calibration}
We first estimate the time-variable instrumental and ionospheric gain, $\mathrm{G_i(t)}$, normalized over all antenna elements of the array. We choose a single frequency ($\nu=\nu_0$) channel for intensity self-calibration and set $\mathrm{B_{i}(\nu_0)}=1$. We write Eq. \ref{eq:decomposed_me} as

\begin{equation}\label{eq:gaincal}
 \begin{split}
 \mathrm{V_{ij}^\prime(\nu_0,\ t)}=&\mathrm{G_{i}(t)\ V_{ij,Gcor}(\nu_0,\ t)\ G_{j}^\dagger(t)}
 \end{split}
\end{equation}
where
\begin{equation}\label{eq:gaincal_1}
\begin{split}
        \mathrm{V_{ij,Gcor}(\nu_0,\ t)}=&\mathrm{B_i(\nu_0)\ K_{\mathrm{cross}}(\nu_0,\ t)\ D_i(\nu_0,\ t)\ V_{ij,app}(\nu_0,\ t)}\\
         &\times \mathrm{D_j^\dagger(\nu_0,\ t)\ K_{\mathrm{cross}}^\dagger(\nu_0,\ t)\ B_j^\dagger(\nu_0)},
\end{split}
\end{equation}
is the apparent model visibility for a single spectral channel after intensity self-calibration.

Our intensity self-calibration algorithm follows the same philosophy as AIRCARS \citep{Mondal2019}, and our implementation incorporates additional improvements and optimizations. Making use of the compact and centrally condensed array configuration of the MWA and the very high flux density of the Sun, we estimate both antenna gains, $\mathrm{G_i(t)}$ and $\mathrm{V_{ij,Gcor}(t)}$ iteratively. 
At any given time stamp, $\mathrm{t}$, we start with the calibration of only the phases of $\mathrm{G_i(t)}$ using all baselines
with at least one of their antennas from the dense central core of the MWA.
Antennas at larger distances are progressively added. The process of intensity self-calibration with increasing number of distant antennas is similar to that implemented in AIRCARS \citep{Mondal2019}. 
Its success has already been demonstrated by many studies relying on high-dynamic range-imaging
\citep[e.g,][]{Mohan2019a,Mohan2019b,Mondal2020a,Mondal2020b,Mohan2021a,Mohan2021b, Kansabanik2022}.

The antennas forming the compact MWA core have the similar ionosphere along their line of sight (LoS) and thus have similar ionospheric phases. 
With the increasing distance from the core, the ionospheric phase for the antenna elements, with respect to a reference antenna in the core, grows. Since the MWA has a centrally condensed core with a large number of antenna elements and a rapid fall off in the density of antennas outside the core, the visibility distribution arising from the core is dominated by the baselines between the core antennas. This provides a certain level of coherency to the baselines originating from the core antennas. A detailed statistical demonstration of this and the advantages it brings is the subject of an independent publication (D. Kansabanik, 2022, in preparation).

The number of baselines originating from the core antennas is more than 70\% of the all available baselines of the MWA. In a nutshell, the coherency of these large number of baselines makes the self-calibration process a well constrained problem, and allows the P-AIRCARS to start the intensity phase-only self-calibration process even without any calibrator source. At initial stages of the intensity self-calibration, the phases of the antenna gains of the farther away antennas are not that well-constrained, as only a small number of baselines involved contribute to the solutions. However, these initial gains are sufficiently close that they provide a good starting point for a self-calibration process. 
Once this process converges, baselines between increasingly distant antennas are included in small steps in subsequent self-calibration rounds.
This is done to ensure that the problem is always well conditioned and most of the antenna solutions are already close to their optimal values. As more longer baselines participate in the calibration, it improves both the gain solutions for the antennas farther away from the core and the source model improve. This process is continued until all of the baselines are included and the final self-calibration iterations have converged. Once all antennas have been added and the phase-only self-calibration has converged, the phases of all antennas are deemed to be determined. Then the algorithm moves to amplitude and phase self-calibration. Once amplitude-phase self-calibration using all the baselines has converged, gains for all the antennas are available. These gain solutions are applied and move to the next steps of the calibration process.

AIRCARS assumes $\mathrm{V_{ij,Gcor}(\nu_0,\ t)}$s to be unpolarized and effectively uses the same source model for the $\mathrm{XX}$ and $\mathrm{YY}$ polarizations.
Unlike AIRCARS, no assumptions are made about the polarimetric properties of $\mathrm{V_{ij,Gcor}(\nu_0, t)}$. 
A $2\times2$ matrix calibration is performed without any  constraints on $\mathrm{V_{ij,Gcor}(\nu_0, t)}$ except that $\mathrm{G_i(t)}$ is assumed to be diagonal.

Figure \ref{fig:intensity_bp_selfcal}a shows the image after the first round of phase-only self-calibration, where the solar disk is rather distorted. The dynamic range of this image is only 24. The image after the first round of amplitude-phase self-calibration is shown in Fig. \ref{fig:intensity_bp_selfcal}b. The coherence of the array has improved remarkably, the solar disk is well formed, and the dynamic range has increased to 385. Figure \ref{fig:intensity_bp_selfcal}c shows the final output of the intensity self-calibration process. The improvement in image quality is self-evident, and the dynamic range has reached to 491. We note that prior to AIRCARS, the highest imaging dynamic range for solar imaging at meter wavelengths was a few hundred ($\lesssim$ 300) and the imaging fidelity was too poor to be able to reliably detect features of strength few percent of the peak \citep{Mercier2009}. 
We have chosen a quiet featureless Sun for this illustration,
the high-dynamic-range and high-fidelity imaging of which continue to remain challenging at low radio frequencies even for the new generation instruments \citep{Vocks2020}.
The presence of solar activity makes the calibration and imaging problem easier, and dynamic ranges exceeding 10$^5$ have been achieved using AIRCARS \citep{Mondal2019}.

\subsection{Bandpass self-calibration}
\begin{figure*}[!t]
    \centering
    \includegraphics[trim={1.5cm 11cm 2cm 1cm},clip,scale=0.95]{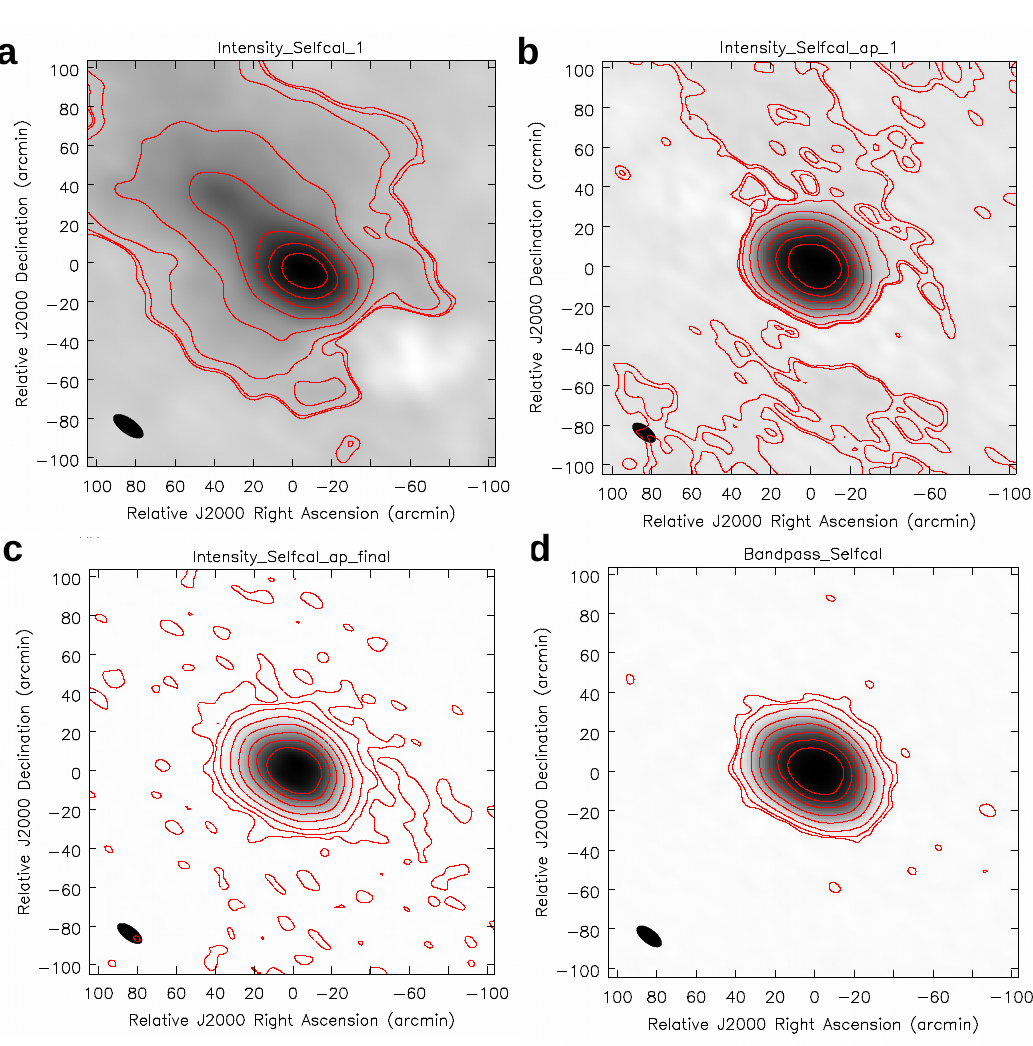}
    \caption{{\textbf{Improvements in image quality during the intensity and bandpass self-calibration.}} Images shown here are at 88 $\mathrm{MHz}$ with a frequency resolution of 160 $\mathrm{kHz}$ and temporal resolution of 2 $\mathrm{s}$. The red contours levels are at 0.3, 0.6, 2, 8, 20, 40, 60, 80 \% of the peak flux density. The black ellipse at the bottom left is the point spread function. a. Image after the first round of intensity self-calibration. Bright emission is present near the phase center, but the solar disk is distorted. b. Image of the first amplitude-phase intensity self-calibration. c. Image after the end of the amplitude-phase intensity self-calibration. The noise level decreases significantly in comparison. d. Image showing the effect of bandpass self-calibration. The image noise is reduced further after the bandpass self-calibration. The dynamic ranges of the images are 24, 385, 491, and 793, respectively.}
    \label{fig:intensity_bp_selfcal}
\end{figure*}

As the instrumental bandpass amplitude and phase vary across frequency, bandpass calibration is required before combining multiple spectral channels to make an image. 
As AIRCARS was designed for spectroscopic imaging  and the flux density calibration was done using an independent nonimaging technique \citep{oberoi2017}, it did not need to include bandpass calibration. Conventionally, instrumental bandpass is determined using standard flux density calibrator sources with known spectra.
Lack of availability of suitable calibrators pushes us to rely on bandpass self-calibration. This in turn required us to find a way to deal with the degeneracy between the instrumental bandpass shape and the intrinsic spectral structure of the source (Sun). To avoid intrinsic spectral structure, we carefully choose sufficiently quiet times for the bandpass self-calibration from the initial flux-density-calibrated dynamic spectrum. 
We obtain the initial flux density calibrated dynamic spectrum using the nonimaging technique mentioned earlier, which is independent of instrumental gains and is computationally much faster.
Though precise flux density calibration of MWA solar observations can be done using the method described in \citet{Kansabanik2022}, it is only applicable to bandpass calibrated visibilities or images, which are not yet available at this stage in the calibration process.
Bandpass self-calibration is performed for narrow bandwidths of 1.28 $\mathrm{MHz}$ at a time, referred to as a `picket'.
The spectrum of the quiet Sun can justifiably be assumed to be flat across a picket and apparent source visibility after bandpass calibration, $\mathrm{V_{ij,Bcor}(\nu,\ t)=V_{ij,Bcor}(\nu_0,\ t)}$.

A single time slice ($\mathrm{t=t_0}$) is chosen for bandpass self-calibration. Bandpass self-calibration starts with $\mathrm{V_{ij,Gcor}(\nu_0,\ t_0)}$ as the initial model. Equation \ref{eq:gaincal_1} over the band can then be rewritten as,

\begin{equation}\label{eq:bandpass}
\begin{split}
 \mathrm{V_{ij,Gcor}(\nu,\ t_0)}=& \mathrm{B_i(\nu)\ V_{ij,Bcor}(\nu_0,\ t_0)\ B_j^\dagger(\nu)}
 \end{split}
\end{equation}
where $\mathrm{V_{ij}(\nu,\ t_0)=V_{ij,Bcor}(\nu_0, t_0)}$ is the apparent model visibility after bandpass self-calibration given as,
\begin{equation}\label{eq:bandpass_1}
\begin{split}
        \mathrm{V_{ij,Bcor}(\nu,\ t_0)}=&\mathrm{V_{ij,Bcor}(\nu_0, t_0)=K_{\mathrm{cross}}(\nu,\ t_0)\ D_i(\nu,\ t_0)}\\
        &\times \mathrm{V_{ij,app}(\nu_0,\ t_0)\ D_j^\dagger(\nu,\ t_0)\ K_{\mathrm{cross}}^\dagger(\nu,\ t_0)}
\end{split}
\end{equation}
We find that inter-picket bandpass phases for MWA can be modeled well by a straight line \citep{Sokolwski2020}. Inter-picket bandpass amplitudes show a more complicated variation and are corrected using an independent method described by \cite{Kansabanik2022}. As expected, bandpass self-calibration improves the image quality, and the dynamic range increases from 431 (Fig. \ref{fig:intensity_bp_selfcal}c) to 793 (Fig. \ref{fig:intensity_bp_selfcal}d).

\subsection{Polarization self-calibration}\label{subsec:polselfcal}
This section describes the different parts of the polarization self-calibration algorithm marked by the orange shaded region in Fig. \ref{fig:polcal_flowchart}.

\subsubsection{Cross-hand phase calibration}
\label{subsec:cross-hand-phase}
During the intensity and bandpass self-calibration, the phase of the reference antenna is set to zero for each of the orthogonal receptors. There is, however, an arbitrary phase difference between the two orthogonal receptors of the reference antenna. This cross-hand phase, $\psi$, is thus a single number and is applied to all antennas by a single Jones matrix $\mathrm{K_{cross}}(\nu,\ t)$. In case of linearly polarized receptors, $\mathrm{K_{cross}}(\nu,\ t)$ causes a leakage from Stokes U to Stokes V and vice versa. 

\begin{figure*}[!t]
 \centering
 \includegraphics[trim={1.3cm 3cm 1.5cm 0cm},clip,scale=0.67]{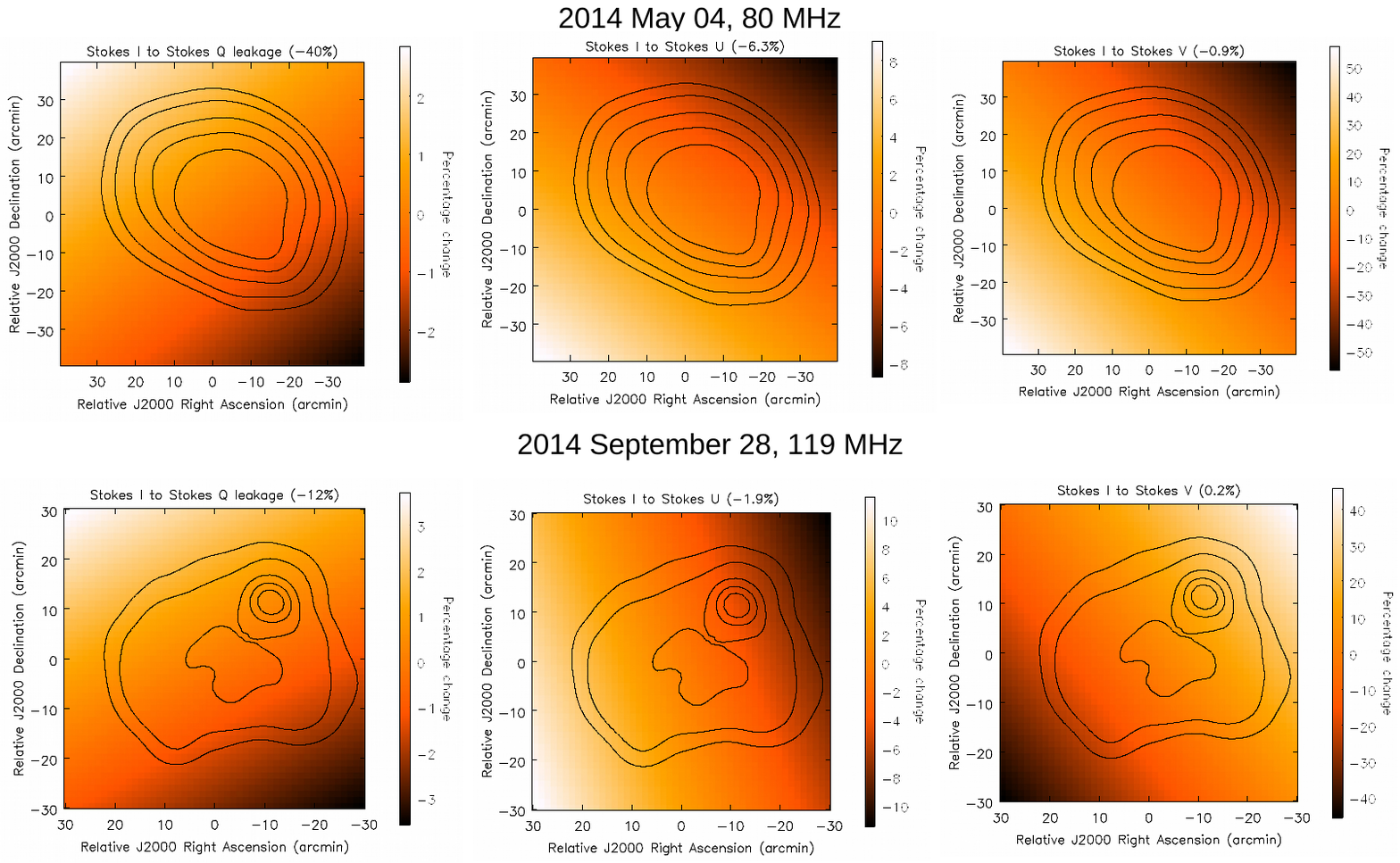}
 \caption{\textbf{Fractional change of leakages from Stokes I to other Stokes due to ideal primary beam response, $\mathrm{E_i(\vec{l})}$, over the angular extend of the Sun.} Black contour represents the Stokes I solar emission. Contour levels are at 10\%, 20\%, 40\%, 60\%, 80 \% of the peak flux density. Polarization leakage from Stokes at the  center of the Sun is mentioned at the title of each panel. \textbf{Top row:} Variation of leakages for the observation of 2014 May 04 at 80 MHz are shown. \textbf{Bottom row:} Variation of leakages for the observation of 2014 September 28 at 119 MHz are shown.}
 \label{fig:stokes_leakage}
\end{figure*}

There are, however, only a few bright polarized sources available for cross-hand phase calibration at low frequencies \citep{lenc2017,lenc2018}.  
In addition to the reasons given in Sec. \ref{sec:conventional-algos}, the requirement of a strong polarized point source at the phase center dominating the measured visibility, imply that the standard cross-hand phase calibration method available in {\sc CASA} cannot be used for aperture arrays with large FoVs.
Cross-hand phase cannot be determined using the self-calibration-based approaches. 
We use an image-based method for calibration of $\psi$, tailored for low-frequency aperture array instruments with large FoV \citep{bernardi2013}.  
An observation of a linearly polarized source with sufficiently high rotation measure (RM) is chosen and the direction-independent instrumental gain and bandpass calibrations obtained from suitable unpolarized calibrator observation are applied. It is important to distinguish the true source polarization from the leakage from Stokes I due to the instrumental primary beam. 
The ideal primary beam corrections are hence applied to account for instrumental leakage.
Even after the ideal primary beam correction, the  residual leakages from Stokes I to other Stokes parameters still remain. For well-designed and well-modeled instruments, departures from nonorthogonality of the dipoles are small, which implies that the Stokes Q to Stokes V leakage is small. For the MWA it is small enough to be ignored \citep{bernardi2013,lenc2017}. 
When a linearly polarized emission passes through magnetized plasma, the polarization angle ($\chi$) rotates. The observed $\chi \propto \lambda^2$, where $\lambda$ is the wavelength of observation. The proportionality constant is referred to as Rotation Measure (RM) and, like all propagation effects, is an integral along the entire LoS. It depends on the distributions of the LoS component of magnetic field strength and electron density. Multiple mediums (e.g., interstellar medium, interplanetary medium and ionosphere) with different RMs contribute to the total rotation. RM synthesis \citep{Brentjens2005} is Fourier synthesis technique to separate out these RM components in the Fourier domain.
The leakage flux from Stokes I to V can also be thought of as yet another medium contributing to the observed RM. 
This leakage flux
must appear at the instrumental RM in Fourier domain, which is typically a few $\mathrm{rad\ m^{-2}}$. As the linearly polarized source is chosen to be at an RM significantly higher than the instrumental RM, any Stokes V emission detected at source RM in the Fourier space can only arise due to leakages from components that rotate with the source RM. Hence the observed Stokes V emission at source RM must arise due to the leakages from Stokes U. Thus the Stokes U and Stokes V flux density are estimated from the image using RM synthesis.  For linearly polarized receptors, $\mathrm{K_{cross}}(\nu,\ t)$ causes the leakage from Stokes U to Stokes V. We vary $\psi$ between $-180^{\circ}$ and $+180^{\circ}$ and determine the value of $\psi$ for which the spurious Stokes V emission is minimized. 

It has been found that the $\psi$ is extremely stable across both time and frequency for the MWA. 
\citet{lenc2017} found that $\psi$ essentially remains constant across the MWA frequency band from 80 $-$ 300 $\mathrm{MHz}$. The GaLactic and Extragalactic All-sky MWA (GLEAM) survey team (private communication with Xiang Zhang, ICRAR) have recently determined that $\psi$ is stable over time scales of years. 
P-AIRCARS generally uses the values of $\psi$ available {\it a priori}, but does provide the flexibility to estimate it from the nearest observations of a linearly polarized source and apply the corresponding correction.
The extreme stability across time and frequency allows the use of nighttime observations from even months away to estimate $\psi$.

After the correction of the cross-hand phase, $\mathrm{V_{ij,Xcor}}$ at any time, $\mathrm{t}$ and any frequency, $\nu$, is obtained from Eq. \ref{eq:bandpass_1} as 
\begin{equation}\label{eq:crossphase_corrrection}
\begin{split}
\mathrm{V_{ij,Xcor}(\nu,\ t)} &=\mathrm{K_{\mathrm{cross}}(\nu,\ t)^{-1}\ V_{ij,Bcor}(\nu,\ t)}\\ &\times \mathrm{K_{\mathrm{cross}}(\nu,\ t)^{-1\dagger}}\\
&= \mathrm{D_i(\nu,\ t)\ V_{ij,app}(\nu,\ t)\ D_j^\dagger(\nu,\ t)}\\
&=\mathrm{D_i(\nu,\ t)}\\&\times \left[ \iint \mathrm{E_{i}(\nu,\ t,\ \Vec{l})}\ \mathbf{B}(\nu,\ t,\ \Vec{l})\ \mathrm{E_{j}^\dagger(\nu,\ t,\ \Vec{l})}\right.\\
&\times \left. \mathrm{e^{-2\pi i(u_{ij}l+v_{ij}m+w_{ij}(n-1))}}\frac{\mathrm{dl\ dm}}{\mathrm{n}} \right]\\
&\times \mathrm{D_j^\dagger(\nu,\ t)}
\end{split}
\end{equation}

\subsubsection{Ideal primary beam correction}\label{sec:ideal_beam}
For aperture array instruments, a significant part of the instrumental leakage comes from the pointing-dependent instrumental primary beam.
In reality, it is rarely feasible to measure the full-Stokes primary beam for aperture arrays.
Hence, an ideal model of the primary beam, obtained using electromagnetic simulations is generally used. There are a few primary beam models available for the MWA -- an analytical beam model \citep{Ord_2010}; an average embedded element (AEE) beam model by \citet{Sutinjo2015}; and full embedded element (FEE) beam models by \citet{oberoi2017} and \citep{Sokowlski2017}. 
These models have steadily improved in their sophistication and performance.
We have used the most recent ideal primary beam model for the MWA \cite{Sokowlski2017} for which the FEE beams have been simulated using the electromagnetic simulation tool, FEKO{\footnote{\url{www.feko.info}}}. We note that P-AIRCARS allows the flexibility to use any primary beam model and will enable us to benefit from improved models as and when they become available. Naturally, the true primary beam response can deviate from the ideal beam model. 

\begin{figure*}[!t]
 \centering
 \includegraphics[trim={1cm 1.5cm 1cm 2.5cm},clip,scale=0.65                     ]{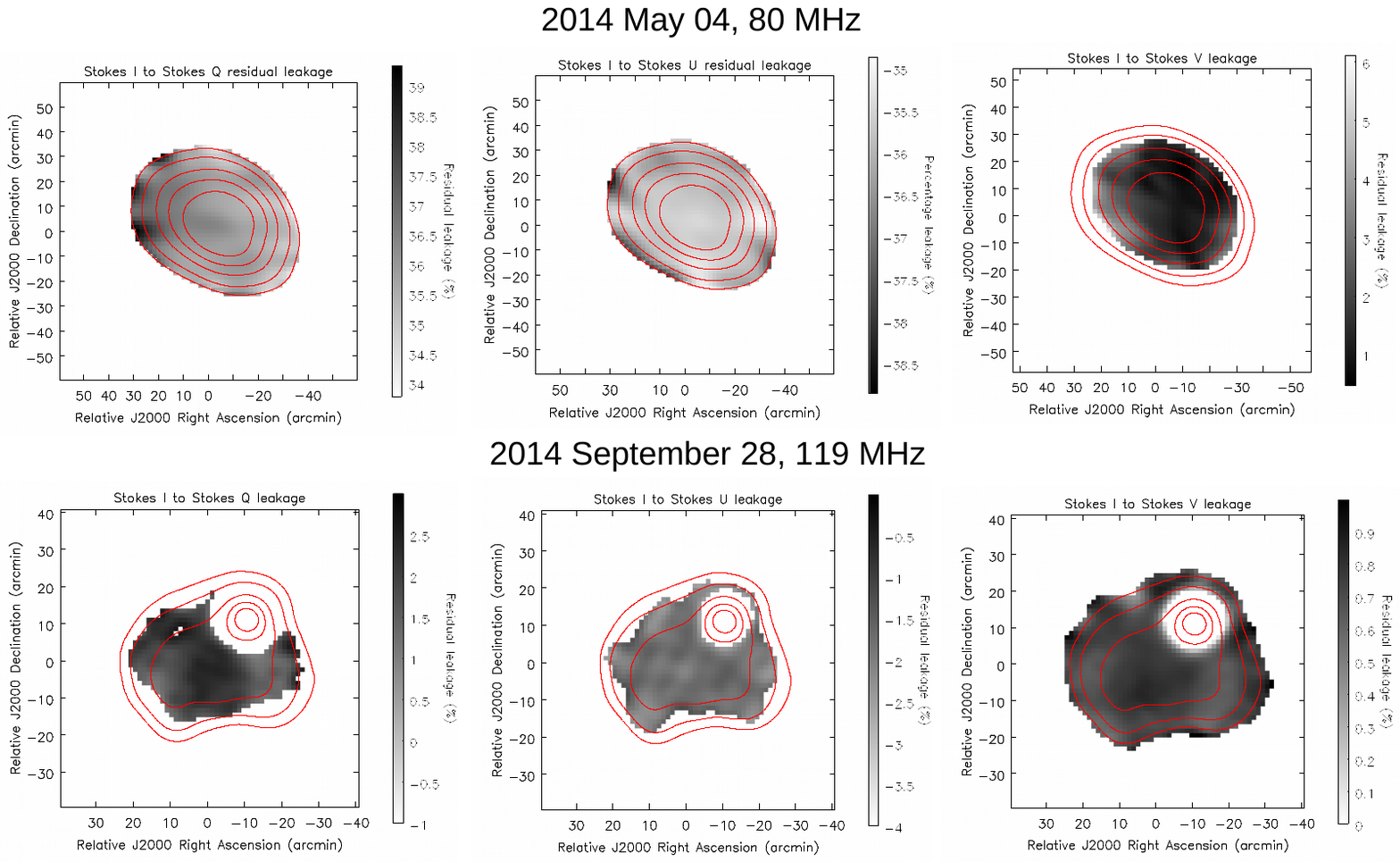}
 \caption{\textbf{Residual leakages from Stokes I to other Stokes over the quiet-Sun region.} Red contour represents the Stokes I solar emission. Contour levels are at 10, 20, 40, 60, 80 \% of the peak flux density. Quiet-Sun regions with more than 3$\sigma$ detection in each Stokes planes are shown. \textbf{Top row: } Residual leakages for the observation on 2014 May 04 at 80 MHz are shown. The beam pointing is at the lowest elevation for the MWA and residual leakages are large. \textbf{Bottom row: }Residual leakages for the observation on 2014 September 09 at 119 MHz are shown. The primary beam pointing is close to meridian, and residual leakages are small. In the Stokes V image for this epoch, there were no quiet-Sun region having more than $3\sigma$ detection. Thus we have only used the $3\sigma$ threshold on Stokes I emission only.}
 \label{fig:res_stokes_leakage}
\end{figure*}

Method I assumes that the leakage from Stokes I to other Stokes components remains essentially constant across the angular span of the solar disk. While this assumption is valid for some pointings, it is not true in general. Percentage variation of this leakage across the solar disk is estimated to be as large as 50\%. We define the percentage variation as,
\begin{equation}\label{eq:frac_leakage_change}
 \begin{split}
 \mathrm{\Delta l(\theta,\ \phi)} = \frac{\mathrm{l(\theta,\ \phi)-l_{centre}}}{\mathrm{l_{center}}}\times100 \%,
 \end{split}
\end{equation}
where $\mathrm{l(\theta,\ \phi)}$ is the leakage from Stokes I at the sky coordinate $(\theta,\ \phi)$, $\mathrm{l_{center}}$ is the leakage at the center of the solar disk, and $\Delta \mathrm{l(\theta,\ \phi)}$ is the percentage change of $\mathrm{l(\theta,\ \phi)}$ with respect to $\mathrm{l_{center}}$. 
The variation of $\Delta \mathrm{l(\theta,\ \phi)}$ over the Sun for two observing epochs is shown in Fig. \ref{fig:stokes_leakage}. 
The Sun was close to half power point of the primary beam during the first epoch, 2014 May 04.
During the second epoch, 2014 September 28, the Sun was close to the peak of the primary beam.
We have found that $\mathrm{l_{center}}$ is smaller for second epoch as compared to that for the first epoch.
This is expected as the primary beam model is more accurate for near its peak. 
Nonetheless, the fractional variation of the leakages over the Sun for both the epochs are similar and not negligible. Hence, for precise polarization calibration, it essential to correct for the direction-dependent primary beam response.

For an homogeneous array comprising identical antenna elements, the the modeled primary beam response can be assumed to be identical for all antenna elements. Thus we can substitute $\mathrm{E_i(\nu,t,\vec{l})=E_j(\nu,t,\vec{l})=E(\nu,t,\vec{l})}$ in Eq. \ref{eq:crossphase_corrrection},
\begin{equation}\label{eq:beamcor_1}
\begin{split}
     \mathrm{V_{ij,Xcor}(\nu,\ t)}&=\mathrm{D_i(\nu,\ t)}\\
     &\times \left[\iint \mathrm{E(\nu,\ t,\ \Vec{l})}\ \mathbf{B}(\nu,\ t,\ \Vec{l})\ \mathrm{E^\dagger(\nu,\ t,\ \Vec{l})}\right.\\
 &\times \left. \mathrm{e^{-2\pi i(u_{ij}l+v_{ij}m+w_{ij}(n-1))}}\frac{\mathrm{dl\ dm}}{\mathrm{n}}\right]\\
 &\times \mathrm{D_j^\dagger(\nu,\ t)}
\end{split}
\end{equation}
When a calibrator source is available, $\mathbf{B}(\nu,\ t,\ \Vec{l})$ is known and the only unknowns in Eq. \ref{eq:beamcor_1} are the $\mathrm{D_i(\nu, t)}$ s. 
As no suitable calibrator observation is available, $\mathrm{D_i(\nu,t)}$ is not known a priori there is a degeneracy between $\mathbf{B}(\nu,\ t,\ \Vec{l})$ and $\mathrm{D_i(\nu,t)}$.
A perturbative approach is used to break this degeneracy.
As $\mathrm{D_i(\nu,\ t)}$ is small compared to $\mathrm{E(\nu,\ t,\ \vec{l})}$s, we approximate $\mathrm{D_i}$ as identity matrix in Eq. \ref{eq:beamcor_1} and obtain the source visibility, $\mathbf{B}_0(\nu,\ t,\ \vec{l})$,
while incorporating the correction for $E(\vec{l})$. 
Equation \ref{eq:beamcor_1} is then takes the form,
\begin{equation}\label{eq:beamcor_2}
\begin{split}
     \mathrm{V_{ij,Xcor}(\nu,\ t)}&=\iint \mathrm{E(\nu,\ t,\ \Vec{l})\ \mathbf{B}_0(\nu,\ t,\ \Vec{l})\ E^\dagger(\nu,\ t,\ \Vec{l})}\\
 &\times \mathrm{e^{-2\pi i(u_{ij}l+v_{ij}m+w_{ij}(n-1))}}\frac{\mathrm{dl\ dm}}{\mathrm{n}}\\
    &=\iint \mathbf{B}_{0,app}(\nu,\ t,\ \Vec{l})\\
    &\times \mathrm{e^{-2\pi i(u_{ij}l+v_{ij}m+w_{ij}(n-1))}}\frac{\mathrm{dl\ dm}}{\mathrm{n}},
\end{split}
\end{equation}
where $\mathbf{B}_{0,app}(\nu,\ t,\ \vec{l})$ is the apparent source visibility, and $\mathbf{B}_0(\nu,\ t,\ \vec{l})$ is the source visibility without the correction for $\mathrm{D_i(\nu, t)}$s and $\mathrm{E(\nu, t, \vec{l})}$. $\mathbf{B}_0(\nu,\ t,\ \vec{l})$ and $\mathbf{B}_{0,app}(\nu,\ t,\ \vec{l})$ are related as
\begin{equation}\label{eq:beamcor_3}
    \begin{split}
        \mathbf{B}_0(\nu,\ t,\ \vec{l})=\mathrm{E^{-1}(\nu,\ t,\ \vec{l})} \mathbf{B}_{0,app}(\nu,\ t,\ \vec{l})\mathrm{E^{-1\dagger}(\nu,\ t,\ \vec{l})}.
    \end{split}
\end{equation}
$\mathbf{B}_0(\nu,\ t,\ \vec{l})$ differs from the true brightness matrix, $\mathbf{B}(\nu,\ t,\ \vec{l})$, due to the following two reasons:
\begin{enumerate}
    \item $\mathrm{D_i(\nu, t)}$ has been ignored in Eq. \ref{eq:beamcor_1}, which introduces errors in $\mathbf{B}_0(\nu,\ t,\ \vec{l})$.
    \item In absence of calibrator observations or other independent astronomical constraints, the degeneracy between $\mathrm{G_i}$ and $\mathrm{V_{ij,Gcor}}$ cannot be broken unambiguously.
\end{enumerate}

Although $\mathrm{D_i(\nu, t)}$ is expected to be small, this may not hold true for all pointings,  especially the ones at low elevations. 
The top row of Fig. \ref{fig:res_stokes_leakage} shows the residual leakages after the ideal primary beam correction at 80 MHz for 2014 May 04. 
For this epoch the beam pointing was at the lowest permitted elevation for the MWA. 
The residual leakages for Stokes I to Stokes Q and U are $\sim34-35\%$ and Stokes I to Stokes V is $\sim3-4\%$. 
For comparison, the bottom row of Fig. \ref{fig:res_stokes_leakage} shows the residual leakages for observation on 2014 September 28 at 119 MHz. 
The beam pointing for this epoch was close to the meridian, and the residual leakages are only a few percent. No systematic spatial variation of the residual leakages is seen in Fig. \ref{fig:res_stokes_leakage}. 
The spatial variations over small angular scales are at 1--2\%  
and are discussed in detail in Sec. \ref{subsec:imaged_based_cor}.
For precise polarization calibration, it is essential to reliably correct for $\mathrm{D_i(\nu, t)}$s.
The next section describes our approach for an image-based first-order correction for $\mathrm{D_i(\nu, t)}$ s to $\mathbf{B}_0(\nu,\ t,\ \vec{l})$.

\subsection{Image-based Leakage correction}\label{subsec:imaged_based_cor}
As discussed in Sec. \ref{sec : basic_polarimetry}, the objective of polarimetric calibration is to correct for instrumental {\it polconversion} and {\it polrotation}, which cannot be achieved using self-calibration-based approaches alone. 
The most recent MWA primary beam model \citep{Sokowlski2017} used in P-AIRCARS successfully reduces the instrumental leakages significantly. 
Though it can come close, no primary beam model can reproduce the true response exactly.
To remove the residual instrumental leakages, 
we use some well-established physical properties of the quiet-Sun thermal emission at metee wavelengths to design an image-based correction. 
The characteristics of the quiet-Sun thermal emission we rely on are:

\begin{enumerate}
\item The brightness temperature of the quiet-Sun thermal emission is well known to lie in the range of $10^5-10^6\ \mathrm{K}$ \citep{Mercier2005, oberoi2017, Vocks2020, Mondal2019, Sharma2020}.
\item Quiet-Sun thermal emission has a very low level of circular polarization ($\lesssim1\%$), which arises due to propagation effects through the magnetized corona \citep{Sastry_2009}.
\item No linearly polarized emission is expected from the quiet-Sun thermal emission \citep{Alissandrakis2021}. 
\end{enumerate}
No assumptions are made about the polarization properties of any active emissions as they depend on the emission mechanism, magnetic field strength, and topology, and are variable across time and frequency. 

Given these properties, one can argue that any Stokes Q and U emission seen in the quiet Sun region must arise due to residual instrumental leakages. 
Also, as the circular polarization of the quiet-Sun is $\lesssim1\%$, the dominant contribution to the Stokes Q and U leakage must come from Stokes I.
For the MWA, the misalignment of the dipoles with respect to the sky coordinates has been established to be small enough to give rise to  insignificant mixing between Stokes Q and U \citep{lenc2017}.
An explicit correction for $\mathrm{K_{cross}}(\nu,\ t)$ has also been applied to correct for the mixing between Stokes U and V.
Based on these arguments, instrumental leakages are corrected as follows:
\begin{enumerate}

\item A $\mathrm{T_B}$ map is first made using the flux-density-calibrated solar images \citep{Kansabanik2022}.
\item Regions where solar emission has reliably been detected are identified using an $n\sigma$ lower threshold, where $\sigma$ is the map RMS in a region far from the Sun, and $n$ is usually chosen to lie between 10 and 6. This is the same threshold as used in the `CLEAN'ing process during imaging and is determined and applied for each Stokes plane independently.
\item The regions lying between $10^5$ and $10^6\ \mathrm{K}$ are considered to correspond to the quiet Sun.
\item Median values of Stokes Q and U fractions from the quiet-Sun regions are computed and deemed to represent the leakages from Stokes I to Stokes Q and U, respectively.
\item The leakages thus determined are then subtracted from the Stokes Q and U maps of $\mathbf{B}_0(\nu,\ t,\ \vec{l})$.

\end{enumerate}
We have only considered the leakages from Stokes I to Stokes Q and U.
Stokes I to V leakages have generally been found to be consistent with 0\% after the correction using the latest FEE beam model \citep{Sokowlski2017}. 
As discussed in Sec. \ref{subsec:cross-hand-phase} Stokes Q to Stokes V leakages are also small enough to be ignored for the MWA. 
Occasionally, however, when the beam pointing is at very low elevations, residual Stokes I to V leakage can grow to be as large as a few tens of percent. 
In such instances, an approach similar to what is used for estimating the first-order corrections for Stokes Q and U maps is used for obtaining a first-order estimate of Stokes I to V leakage. 
This is employed only when the median circular polarization in the quiet-Sun region is found to be $> 2\%$. 
Method I used an older beam model \citep{Sutinjo2015} which could not correct for the Stokes I to V leakage effectively, and needed to rely exclusively on this approach to estimate Stokes I to V leakage.
In contrast, we use it sparingly, only when the Stokes I to V leakage is so large that our perturbative approach breaks down.
These corrections account for {\it polconversion}, which arises due to the $\mathrm{D_i(\nu, t)}$ terms.
There is little mixing between any of Stokes Q and U or Stokes Q and V for the MWA \citep{bernardi2013,lenc2017}. 
As $\mathrm{K_{cross}}$ has also been corrected for, there is no mixing between Stokes U and V.
Hence, no additional correction for polrotation is required.

For aperture arrays, errors on the ideal beam are expected to be direction-dependent.
This is true for the MWA as well \citep{lenc2017}. 
We find that, although the leakages due to the ideal primary beam varies significantly over the angular extent of the Sun, the residual leakages due to $\mathrm{D_i(\nu, t)}$s do not. 
Figure \ref{fig:res_stokes_leakage} shows the residual leakages from Stokes I to other Stokes parameters after the ideal primary beam correction.
No significant systematic variations are seen across the solar disk after the subtraction of the median leakage, validating the assumption to treat the residual leakage or $\mathrm{D_i(\nu, t)}$s as direction-independent quantities, as mentioned in Sec. \ref{paircars_algorithm}.
This image-based correction for leakages yields the source brightness matrix, $\mathbf{B}_1(\nu,\ t,\ \vec{l})$, which incorporates the first-order corrections for these leakages.
Paralleling  Eq. \ref{eq:beamcor_2}, $\mathbf{B}_{1,app}(\nu,\ t,\ \vec{l})$ can be written as:
\begin{equation}\label{eq:leakge_cor1}
    \begin{split}
        \mathbf{B}_{1,app}(\nu,\ t,\ \vec{l})&=\mathrm{E(\vec{\nu,\ t,\ l})}\ \mathbf{B}_1(\nu,\ t,\ \vec{l})\ \mathrm{E^{\dagger}(\nu,\ t,\ \vec{l})}
    \end{split}
\end{equation}

We can now rewrite Eq. \ref{eq:beamcor_1} using $\mathbf{B}_{1,app}(\vec{l})$ and including the first-order correction of error matrices, $\mathrm{D_{1,i}(\nu, t)}$s, as
\begin{equation}\label{eq:leakge_cor2}
    \begin{split}
        \mathrm{V_{ij,Xcor}(\nu,\ t)}&=\mathrm{D_{1,i}(\nu,\ t)}\
        \left [\iint \mathbf{B}_{1,app}(\nu,\ t,\ \vec{l})\right.\\ &\times \left. \mathrm{e^{-2\pi i(u_{ij}l+v_{ij}m+w_{ij}(n-1))}}\frac{\mathrm{dl\ dm}}{\mathrm{n}}\right]\\
        &\times \mathrm{D_{1,j}^\dagger(\nu,\ t)}\\
        \mathrm{V_{ij,Xcor}(\nu,\ t)}&=\mathrm{D_{1,i}(\nu,\ t)\ V_{1,ij,app}(\nu,\ t)\ D_{1,j}^\dagger(\nu,\ t)},\\
    \end{split}
\end{equation}
where $\mathrm{V_{1,ij,app}(\nu, t)}$ is the apparent source visibility after first-order correction. 
$\mathrm{D_{1,i}(\nu, t)}$s can now be solved for iteratively using $\mathrm{V_{1,ij,app}}$ as the initial model, as discussed next.

\subsubsection{Perturbative Correction : Residual Poldistortion Estimation and Correction}\label{sec:poldist}

The $\mathrm{V_{1,ij,app}(\nu, t)}$s can be thought of as the full-Stokes sky model but with small errors arising largely from the deficiencies of first-order polarization calibration. 
The situation is analogous to the missing flux density problem in intensity self-calibration \citep{Grobler2014}, where it is addressed by using normalized solutions over the all antenna elements of the array.
For the same reasons, a normalization factor also needs to be estimated in case of polarization self-calibration. 

As the first-order corrections have already been applied in the process of determining $\mathrm{V_{1,ij,app}(\nu, t)}$, $\mathrm{D_{1,i}(\nu, t)}$ is expected to be small. $\mathrm{D_{1,i}}$ is estimated using full Jones matrix solver {\it CubiCal}.
We define
\begin{equation}\label{eq:pol_selfcal_1}
 \begin{split}
 \mathrm{D_{1,i}(\nu, t)}&=\mathrm{D_{i}(\nu, t)\ P_D(\nu, t)},
 \end{split}
\end{equation}
where $\mathrm{D_{i}(\nu, t)}$ represents the true values of the errors on the ideal instrumental primary beam mentioned in Eq. \ref{eq:beamcor_1}, and $\mathrm{P_D(\nu, t)}$ is the residual {\it poldistortion} left behind in the data after the first-order corrections.
The mean of all $\mathrm{D_{i}(\nu, t)}$ is expected to lie close to the identity matrix and is the full-Stokes analog of the scalar normalization used in intensity self-calibration.
For ease of notation we drop the explicit $\nu$ and $t$ dependence of $\mathrm{V_{1,ij,app}}$s, $\mathrm{D_{1,i}}$s, $\mathrm{D_{i}}$s and $\mathrm{P_D}$s in the following text.

We initiate the self-calibration process using $\mathrm{V_{1,ij,app}}$s as the initial model. While $\mathrm{D_{1,i}}$ is not assured to be close to the Identity matrix, the first-order calibration already applied makes them small enough for a self-calibration-like approach to converge.
$\mathrm{P_D}$ is estimated assuming that all $\mathrm{D_{i}}$ are close to identity. 
To estimate $\mathrm{P_D}$, $\mathrm{S_D}$, the sum of the variance of $\mathrm{D_{i}}$ with respect to identity matrix, $\mathrm{I}$, is minimized. $\mathrm{S_D}$ is defined as:
\begin{equation*}
\begin{split}
 \mathrm{S_D} & =\mathrm{\sum_ivar(D_\mathrm{i}-I)}\\
 & = \mathrm{\sum_i var(D_{1,i}\ P_D^{-1}-I)}\\
 & = \mathrm{\sum_i Tr \left[ (D_{1,i}\ P_D^{-1}-I)(D_{1,i}\ P_D^{-1}-I)^\dagger \right]}\\
 & = \mathrm{\sum_i Tr \left[D_{1,i}\ P_D^{-1}\ P_D^{\dagger-1}D_{1,i}^\dagger\right]-Tr\left[P_D^{\dagger-1}\ D_{1,i}^\dagger\right]}\\
 & \mathrm{\qquad\qquad -Tr\left[D_{1,i}\ P_D^{-1}\right]+Tr\left[I\right]}\\
\end{split}
\end{equation*}
For minimization, $\frac{\mathrm{\partial S_D}}{\mathrm{\partial P_D}} =0$ is imposed, leading to the following relation:  

\begin{equation}
\begin{split}
\left (\mathrm{\sum_i P_D^{-1} D_{1,i} P_D^{-1}}\right)^\dagger & =\biggl( \mathrm{\sum_i P_D^{-1} P_D^{\dagger-1} D_\mathrm{1,i}^\dagger D_\mathrm{1,i} P_D^{-1} \biggr)^\dagger}\\
\mathrm{P_{D}^{\dagger-1}\left(\sum_iD_{1,i}^\dagger\right)P_{D}^{\dagger-1}} &=  \mathrm{P_{D}^{\dagger-1} \left(\sum_i D_{1,i}^\dagger D_{1,i} P_{D}^{-1}\right) P_{D}^{\dagger-1}}\\
 \left(\mathrm{\sum_i D_{1,i}^\dagger D_{1,i}}\right) \mathrm{P_D^{-1}} & =\mathrm{\sum_i D_{1,i}^\dagger}\\
 \mathrm{P_{D}^{-1}} & =\left(\mathrm{\sum_i D_{1,i}^\dagger D_{1,i}}\right)^{-1} \mathrm{\sum_i D_{1,i}^\dagger}\\
 \mathrm{P_D} & =\left[ \left(\mathrm{\sum_i D_{1,i}^\dagger D_{1,i}}\right)^{-1} \mathrm{\sum_i D_{1,i}^\dagger}\right]^{-1}
\end{split}
\end{equation}

We then correct each of the $\mathrm{D_{1,i}}$ for the $\mathrm{P_D}$ and obtain $\mathrm{D_{i}}$ as given by:
\begin{equation} \label{eq:model-error}
\begin{split}
 \mathrm{D_{i}}&=\mathrm{D_{1,i}\ P_D^{-1}}.
\end{split}
\end{equation}
Using Eq. \ref{eq:model-error}, Eq. \ref{eq:leakge_cor2} can be written as,
\begin{equation}
 \begin{split}
\mathrm{V_{ij,Xcor}}=&\mathrm{D_{1,i}\ V_{1,ij,app}\ D_{1,j}^\dagger}\\
=&\mathrm{D_{i}\ P_D\ V_{1,ij,app}\ P_D^\dagger\ D_{j}^\dagger}\\
=& \mathrm{D_{i}\ V_{ij,app}\ D_{j}^\dagger}\\
\mathrm{V_{ij,app}}=& \mathrm{D_{i}^{-1}\ V_{ij,Xcor}\ D_{j}^{-1\dagger}}, 
 \end{split}
\end{equation}
where $\mathrm{V_{ij,app}}$ is the final apparent source visibility. 
The final true source brightness matrix, $\mathbf{B}(\nu,\ t,\ \vec{l})$, is then obtained using the ideal primary beam as:
\begin{equation}
 \mathbf{B}(\nu,\ t,\ \vec{l})=\mathrm{E(\nu,\ t,\ \vec{l})^{-1}}\ \mathbf{B}_{app}(\nu,\ t,\ \vec{l})\ \mathrm{E(\nu,\ t,\ \vec{l})^{-1\dagger}}.
\end{equation}

\begin{figure*}[!t]
    \centering
    \includegraphics[trim={3.2cm 1.5cm 6.5cm 1cm},clip,scale=0.88]{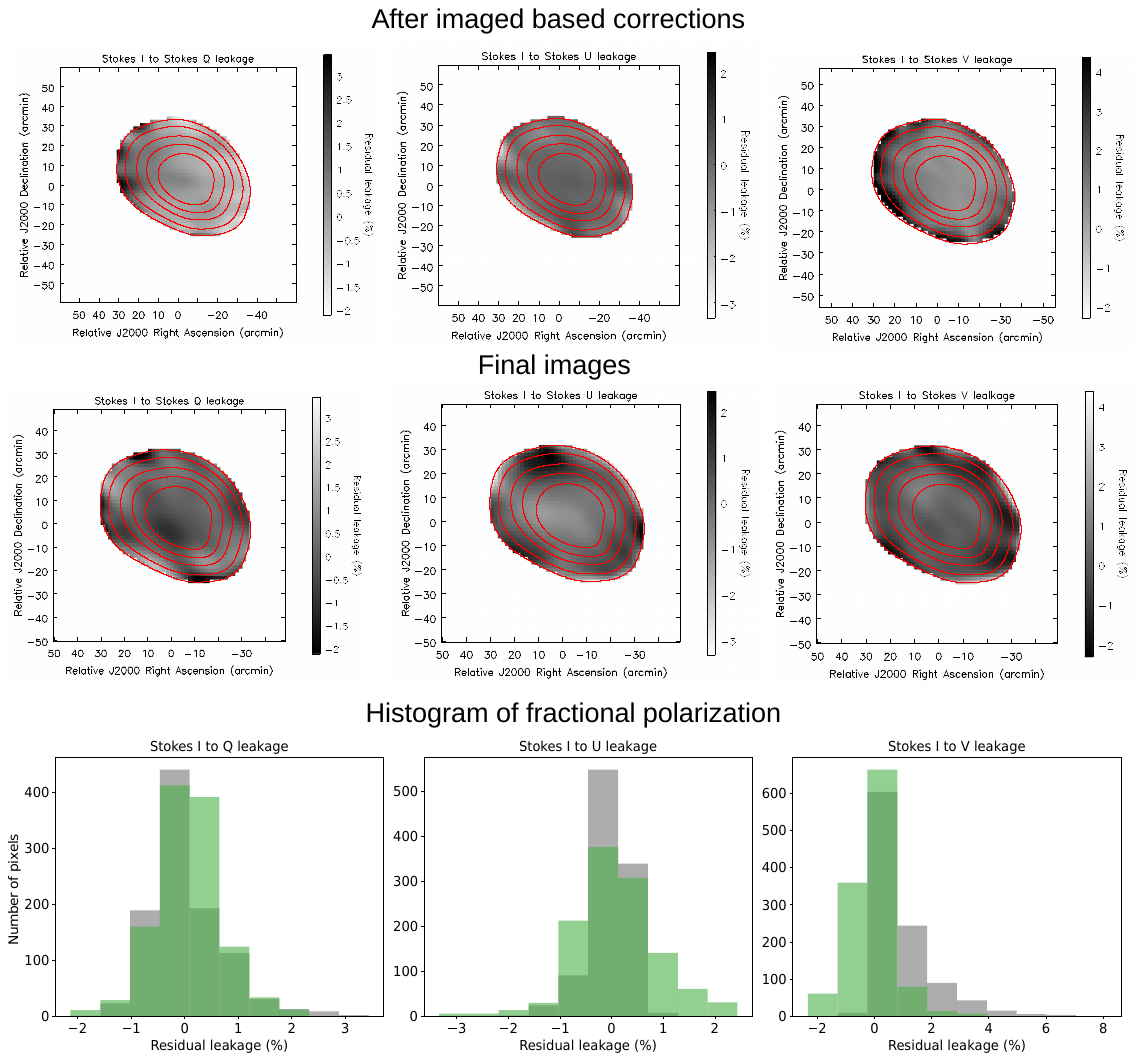}
    \caption{\textbf{Residual Stokes leakages after the image-based correction and in the final images for the observing epoch 2014 May 04 at 80 MHz.} Red contours represent the Stokes I emission. Contour levels are at 10, 20, 40, 60, 80\% of the peal Stokes I flux density. Polarization percentages over the regions with more than 3$\sigma$ detection in Stokes I are shown in gray color scale. The color scale for the corresponding panels in the top and middle rows are identical, and some pixels toward the edge are saturated. The full range of values are shown in the histograms. \textbf{Top row: } Polarization fraction over the quiet-Sun region obtained after the image-based correction are shown. No systematic variation is present. $\mathrm{L_{residual}}$ for Stokes Q, U, and V are respectively $\lesssim3.5,\ 2$ and 1\%. \textbf{Middle row: } Polarization images over the quiet-Sun region after the final self-calibration iteration are shown. $\mathrm{L_{residual}}$ for the Stokes Q, U, and V images are $\lesssim1.8,\ 0.8$ and 0.08\% respectively. \textbf{Bottom row : }Histogram of the pixel values of each Stokes images after the image-based correction are shown in gray in background and of the final images in green in foreground.}
    \label{fig:polcal_progress}
\end{figure*}

\subsection{Estimation of Residual Leakages} 
For astronomical observations, residual leakages in the polarization images are usually determined using unpolarized celestial sources.
The polarized emission detected from such sources, after polarization calibration, provides an estimate of the residual leakages from Stokes I to other Stokes parameters. 
For large FoV instruments, the leakage can be a strong function of direction. Despite this, an approach based on observations of multiple unpolarized sources has been successfully used \citep{lenc2017}.
They first determined the leakages toward the individual sources.
Next a 2D polynomial of second order was fit to determine the leakages as a function of direction.
It has recently been shown that numerous background sources can be detected in Stokes I with the MWA even in presence of the Sun \citep{Kansabanik2022}. 
However, to pursue a similar approach for determining leakages, a large number of these sources needs to be detected in other Stokes parameters as well, which is yet to be demonstrated.
 
We use the quiet-Sun regions in our images to estimate residual leakage fraction, $\mathrm{L_{residual}}$, as follows:
\begin{enumerate}
    \item When Stokes Q, U and V emission is detected with more than $3\sigma$ significance over more than 50\% of the quiet-Sun region, we use the median values of the pixels detected in each Stokes plane, and the define the residual leakages as:
    \begin{equation}\label{eq:median_leakage}
        \mathrm{L_{residual}}=\frac{\mathrm{med(L)_{Q,U,V}}}{\mathrm{I_{max}}},
    \end{equation}
    where $\mathrm{med(L)_Q}$, $\mathrm{med(L)_U}$ and $\mathrm{med(L)_V}$ are the median values of the Stokes Q, U, and V pixels detected with more than $3\sigma$ significance and $\mathrm{I_{max}}$ is the maximum pixel value in $\mathrm{Jy}$ per beam in the quiet-Sun regions.
    \item When no polarization is detected over the quiet-Sun region, we define a residual leakage limit based on the $3\sigma$ limit. In such situations we define the $\mathrm{L_{residual}}$ as:  
    \begin{equation}\label{eq:res_leakage}
  \mathrm{L_{residual}}<\left|\frac{\mathrm{3\times \sigma_{Q,U,V}}}{\mathrm{I_{max}}}\right|,  
\end{equation}
   where $\mathrm{\sigma_Q,\ \sigma_U,\ \sigma_V}$ are the rms noise values in $\mathrm{Jy}$ per beam of the Stokes Q, U, and V images, respectively, close to the Sun and $\mathrm{I_{max}}$ is defined as before.
   \item Quiet-Sun emission in Stokes I may not always be detectable during the presence of very bright radio bursts (e.g., when a type III radio burst is in progress). In such cases, the residual leakage is estimated using the closest time stamp where the Stokes I quiet-Sun emission is detected. 
\end{enumerate}

\subsubsection{A P-AIRCARS Stress Test}
\label{subsec:stress-test}
As a part of the process of development of P-AIRCARS and evaluating its efficacy, it was tested on some particularly challenging datasets.
The most challenging of these observations comes from 2014 May 04 when the MWA was pointed to its lowest permissible elevation, where the sensitivity of the MWA and its polarization response are the poorest. 
This observation was at 80 MHz, where the flux density of the Sun is the lowest and Galactic background the strongest. 
Additionally, these data also correspond to quiet-Sun conditions, where the Sun is essentially a featureless extended source and is hardest to  calibrate and image.
Processing these data, hence, correspond to a stress test of P-AIRCARS.
It is quite reasonable to expect that, if P-AIRCARS can meet the challenge of calibrating and imaging these data, it will be able to successfully deal with most other MWA solar data.
This section substantiates the performance of P-AIRCARS on these data.

The Stokes images after the corrections of the modeled primary beam response are shown in the top row of Fig. \ref{fig:res_stokes_leakage}. $\mathrm{L_{residual}}$ for the Stokes Q, U, and V at this stage is $\sim34,\ -35,$ and $2.8$\% respectively. As the Stokes I to V leakage was more than $2\%$, an imaged-based leakage correction was performed for Stokes V, as discussed in \ref{subsec:imaged_based_cor}. 
The top row of Fig. \ref{fig:polcal_progress} shows the percentage polarization over the quiet-Sun regions with more than $3\sigma$ detection in Stokes I after the image-based leakage corrections.  
The dynamic range of the Stokes I image is $\sim$500, and it is evident that Stokes I to Q, U leakages have reduced by about an order of magnitude (top panel of Fig. \ref{fig:polcal_progress}) beyond what was obtained by primary beam correction (top panel of Fig. \ref{fig:res_stokes_leakage}). In this case Stokes Q emission is detected at $> 3\sigma$ significance over the quiet-Sun region, but that at Stokes U and V is not detected. 
Hence we use Eq. \ref{eq:res_leakage} to obtain the limit of $\mathrm{L_{residual}}$ for Stokes U and V. $\mathrm{L_{residual}}\approx-3.5$ for Stokes Q calculated using Eq. \ref{eq:median_leakage} and $\mathrm{L_{residual}}<|2| $ and $|1|\%$ for Stokes U and V, respectively, are calculated using Eq. \ref{eq:res_leakage}.
No systematic variation of the polarized emission is seen across the solar disk. 
After the image based leakage correction, several rounds of polarization self-calibration are performed and $P_D$ is corrected for at each iteration. This process is deemed converged when the absolute total polarized flux flux densities for Stokes Q, U, and V become stable.
The rms of the Stokes images and the residual leakages are found to improve with every iteration of polarization self-calibration.
The final Stokes Q, U, and V images are shown in the middle row of Fig. \ref{fig:polcal_progress}. Most of the regions in Stokes Q, U and V images are found to be consistent with noise.
The $\mathrm{L_{residual}}$ of the final Stokes Q, U, and V images have reached the values $<|1.8|,\ |0.8|$ and $|0.08| \%$ respectively. 
The histogram of the pixel values of the Stokes Q, U and V leakage fraction are shown in the bottom row of Fig. \ref{fig:polcal_progress}.
The median leakage values are already close to zero after image-based corrections, as expected, but tend to be asymmetric in some cases.
In some cases, low-level artifacts are seen after image-based corrections.
The regions with $>5\sigma$ detection after the image based correction are used for subsequent rounds of polarization self-calibration. The histograms of pixel values for the final images (shown in green) grow more symmetric and demonstrate a reduction in artifacts in the polarization images. 
The magnitude of this improvement can vary across different Stokes planes. 
For example, the histograms for Stokes Q, both after image-based correction (gray) and polarization self-calibration (green), are symmetric and very similar, demonstrating that image-based correction was already very good and did not give rise to any significant artifacts.
On the other hand, after image based correction, the Stokes V histogram shows a very skewed distribution with a positive tail extending out to 7.5\%. 
Polarization self-calibration leads to a much more symmetric distribution with a smaller span of $\pm$2\%.
This demonstrates the efficacy of polarization self-calibration at reducing the artifacts in the Stokes V image. 
The situation for Stokes U is found to lie between these two regimes.
While the improvements are always seen after polarization self-calibration, the trends seen here are not general and can differ from observation to observation.

\subsection{Computational Load}
P-AIRCARS uses the full Jones calibration package {\it CubiCal} \citep{cubical2018,Cubical_robust2019} for calibration and {\it WSClean} \citep{Offringa2014} for imaging. 
The computational load of P-AIRCARS naturally depends on the details of the data, the choices made during analysis and computation hardware available.
We present some numbers here to provide a general sense for the computational load and clock time taken for calibration. 
We use a 2 GHz CPU core with hyperthreading (two threads per core) as the benchmark device.
The very first Stokes I self-calibration run on a chosen time and frequency slice (referred to as the reference time and frequency) takes about an hour. Once the gain solutions obtained from this reference slice have been applied to the dataset, the next step, bandpass self-calibration is performed in parallel on each of the 1.28 MHz wide 24 spectral chunks.
This takes about 15--20 minutes per hyperthreaded core.
The final step of polarization self-calibration takes about 45 minutes per spectral slice and is also parallelized across the frequency axis.
Thus for a typical P-AIRCARS run, full-Stokes calibration takes about 2 hours for a usual MWA dataset with 30.72 MHz bandwidth when parallelized across 25 cores (50 threads).
As calibration (and imaging) are both done in a spectroscopic snapshot mode, only a tiny fraction of the entire dataset is needed for any individual calibration run making the memory footprint very small. 
More details about the implementation of the algorithm, the optimizations used, and the computational load will be presented in a forthcoming paper (D. Kansabanik et al., 2022, in preparation).

\section{Results and Discussion}\label{sec : result}
\begin{figure*}[!t]
    \centering
    \includegraphics[trim={2cm 9cm 3cm 1cm},clip,scale=0.72]{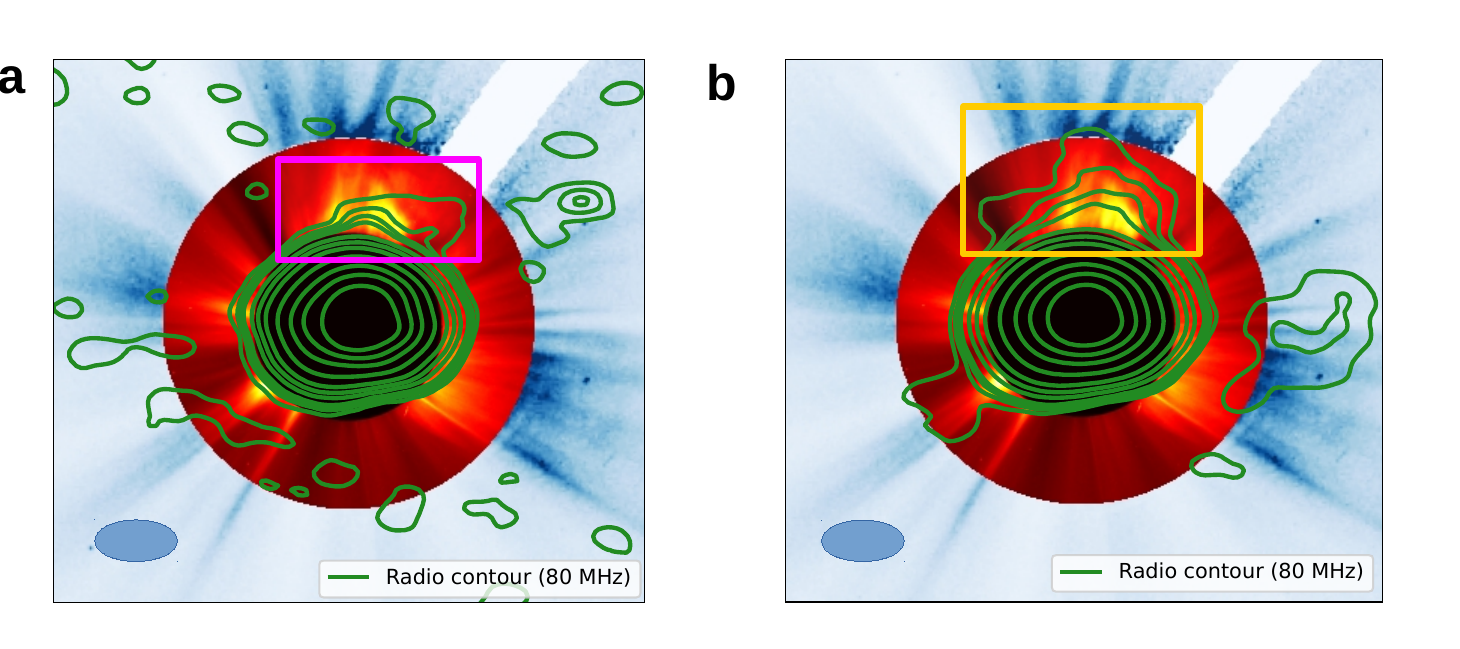}
    \caption{{\textbf{Comparison between images made using AIRCARS and P-AIRCARS.}} The background images are LASCO white light coronagraph images. The red color map represents LASCO C2 images, and blue color map represents the LASCO C3 images. The green contours represents the radio images at 80 MHz. The contour levels are 0.2\%, 0.4\%, 0.6\%, 0.8\%, 2\%, 4\%, 8\%, 20\%, 40\%, 60\%, 80 \% of the peak flux density. The filled ellipses at the lower left of the images are the PSF. {\textbf{a.}} The image made using AIRCARS has small extended emission from the CME as marked by the pink box. There are noise peaks at the 0.2\% level. {\textbf{b.}} The image made using P-AIRCARS has extended emission over a larger region as marked by the yellow box. The radio emission covers the full white light CME.}
    \label{fig:aircars_paircars}
\end{figure*}

\begin{figure*}[!t]
    \centering
    \includegraphics[trim={1.8cm 6.9cm 1.8cm 1cm},clip,scale=0.68]{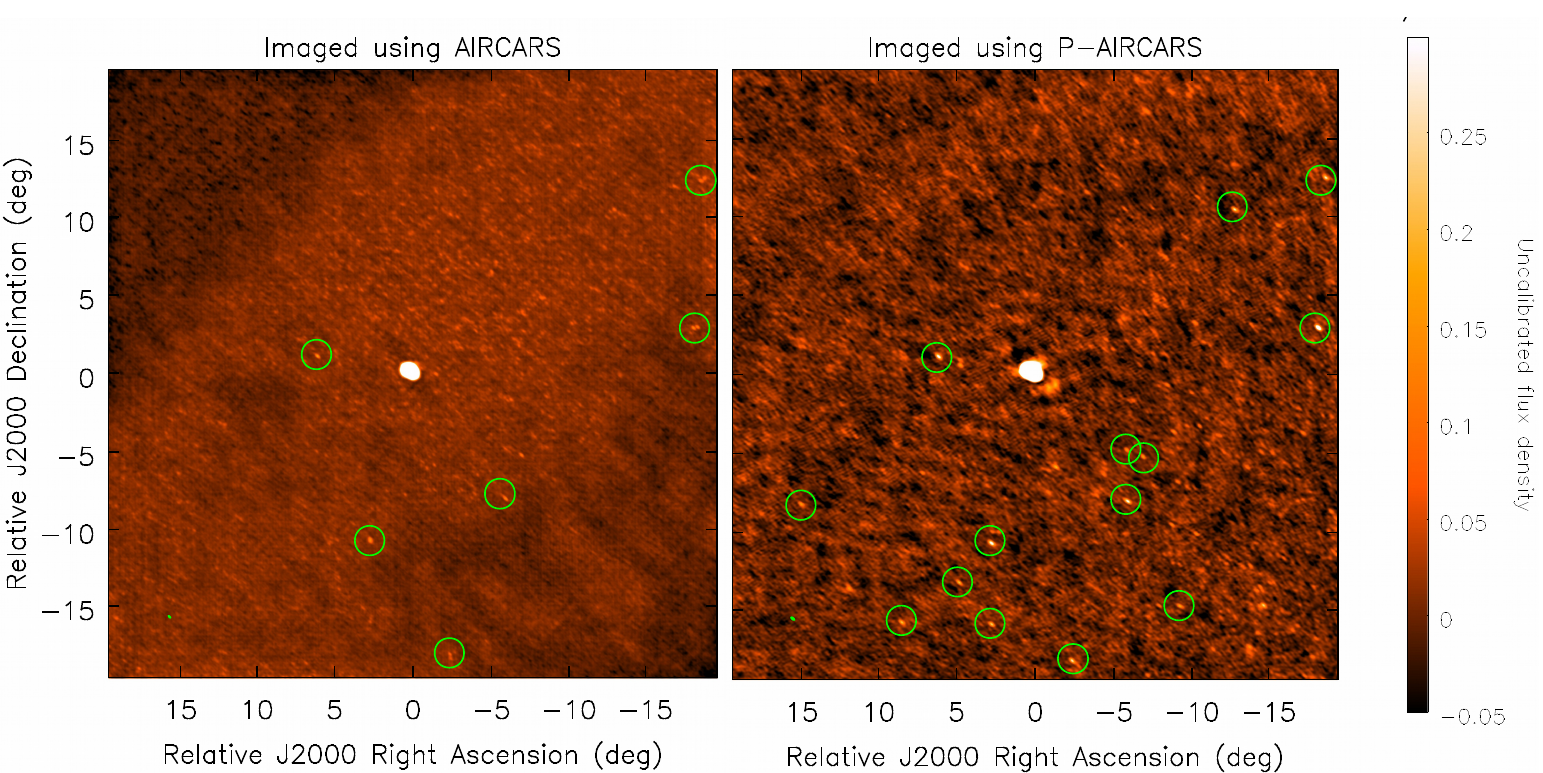}
    \caption{{\textbf{Comparison of the noise characteristics of the images made using AIRCARS and P-AIRCARS.}} The noise characteristic of the same image, as shown in Fig. \ref{fig:aircars_paircars} is shown for the 30$^{\circ}\times$30$^{\circ}$ FoV. The detected background galactic and extragalactic radio sources are marked by green circles. 
    {\textbf{Left panel: }} The AIRCARS image shows a non-uniform noise behavior. A bright wide strip of enhanced noise runs across the full image passing from southwest to the northeast. The dynamic range of this image is $\sim1000$, and 6 background sources are clearly detected. {\textbf{Right panel: }} The P-AIRCARS image show a much more uniform noise characteristic over the full FoV. Both small and large angular scale artifacts have reduced in strength. The dynamic range of the image is $\sim1800$. The substantially improved dynamic range leads to the detection of 14 background sources with higher detection significance in the image.}
    \label{fig:aircars_paircars_rms}
\end{figure*}

The performance of P-AIRCARS on a particularly challenging quiet-Sun dataset has already been substantiated in Sec. \ref{subsec:stress-test}.
This section substantiates the various improvements that P-AIRCARS images represent and the science opportunities they enable.  All of the solar data used here were taken under the MWA Project ID G0002.
P-AIRCARS usually achieves $<|2|\%$ residual leakages for Stokes Q and Stokes U and $<|1|\%$ residual leakage for Stokes V. 
These values are comparable to what is generally achieved for high-quality astronomical observations \citep{lenc2017,lenc2018,Risley2018,Risley2020} with the MWA and much better than those achieved by earlier spectropolarimetric solar studies \citep{Patrick2019,Rahman2020}. 
P-AIRCARS not only delivers a very small residual leakage, it also provides improved Stokes I imaging fidelity as compared to AIRCARS, as shown next.

\subsection{Improvements in Stokes I imaging}
In addition to the operational efficiency and polarimetric imaging capability, P-AIRCARS includes multiple improvements over the earlier Stokes I state-of-the-art pipeline \citep[AIRCARS;][]{Mondal2019}, which was focused on spectroscopic snapshot imaging. It is well unknown that the lack of calibration of instrumental polarization leakages leads to a reduction in the dynamic range \citep{Bhatnagar2001}. Besides that, while imaging over multiple frequency channels, if the instrumental bandpass is not corrected, it leads to artifacts in the images. 
P-AIRCARS includes both these capabilities, which lead to a significant improvement in Stokes I image fidelity and noise properties.

A comparison of radio maps from AIRCARS (left panel) and P-AIRCARS (right panel) made using exactly the same data is shown in Fig. \ref{fig:aircars_paircars}.
The radio maps at 80 MHz have been superposed on  Large Angle and Spectrometric Coronagraph \citep[LASCO][]{Brueckner1995} images and use data of duration of 2 minutes and a bandwidth of 2 MHz.
AIRCARS image shows a small weak emission feature from a CME on 2014 May 04 (marked by the pink box) over a region comparable to the point spread function (PSF). The P-AIRCARS image shows emission at the similar strength but extended over a much larger region covering the full extent of the white light CME (marked by yellow box). 
It is evident that the imaging artifacts near the Sun in the P-AIRCARS image are at a lower level, and another weak extended emission feature lying on the western limb overlapping with the LASCO C3 FoV has a reliable detection. 
The noise characteristics of the P-AIRCARS image have also improved significantly. 
To substantiate this, Fig. \ref{fig:aircars_paircars_rms} shows the entire FoV of the image shown in Fig. \ref{fig:aircars_paircars}. 
It is evident that the noise characteristics of the AIRCARS image (left panel Fig. \ref{fig:aircars_paircars_rms}) is not uniform across the image, there is a bright and extended noise band running diagonally across the image from southwest to northeast. 
Dynamic range of this image is $\sim1000$. 
On the other hand, the noise characteristic of the P-AIRCARS image (right panel of Fig. \ref{fig:aircars_paircars_rms}) is quite uniform, and the large angular scale feature seen in AIRCARS image is no longer evident. 
The dynamic range of the P-AIRCARS image has improved by $\sim80\%$, which leads to the value $\sim1800$. 
\begin{figure*}[!t]
    \centering
    \includegraphics[trim={2cm 9cm 1cm 1cm},clip,scale=0.67]{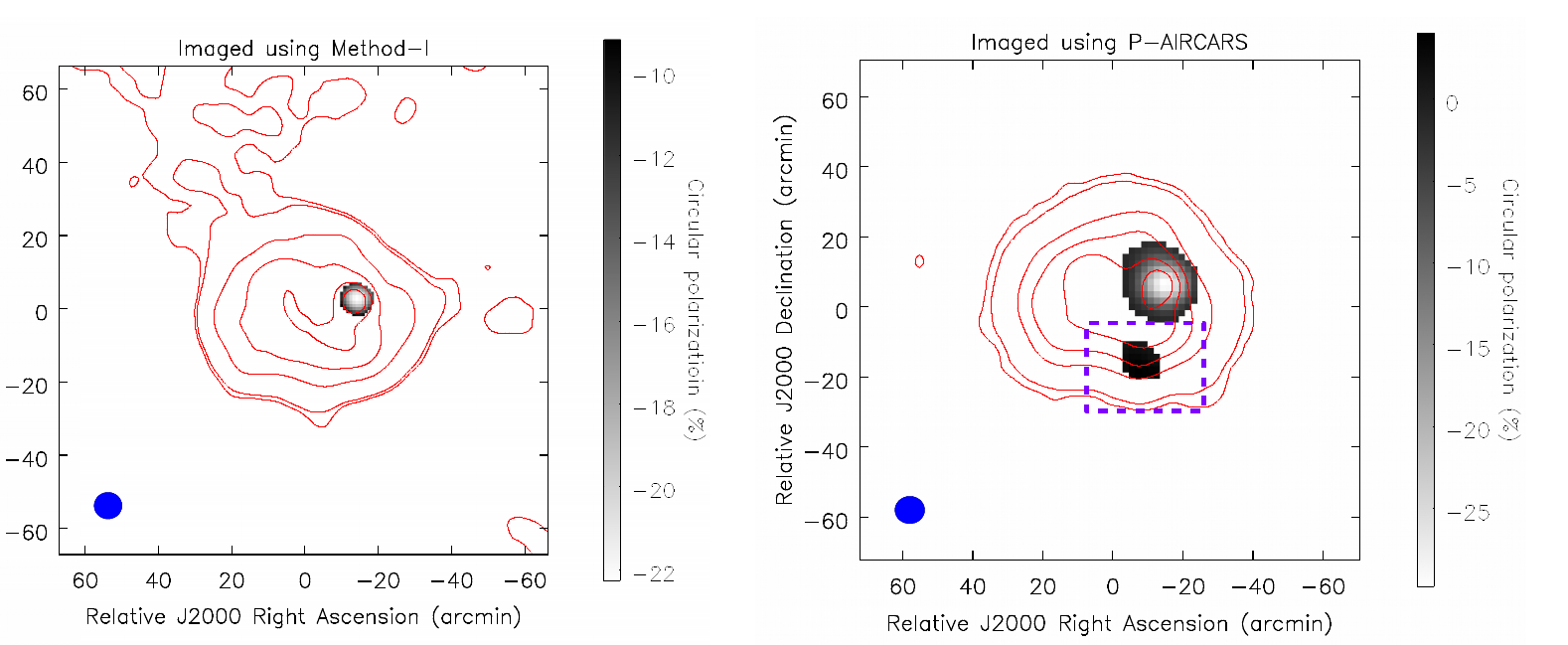}
    \caption{{\textbf{Comparison between Stokes I and Stokes V images made using method I and P-AIRCARS.}} The Stokes I and Stokes V images of an active solar emission observed at 159 MHz at 2014 October 24, 03:46:00 UTC are shown. Circular polarization fraction are shown by gray color map. Only the Stokes V emission detected at more than 3$\sigma$ detection is shown. The red contours represents the Stokes I emission. The contour levels are at 0.8\%, 2\%, 20\%, 40\%, 60\%, and 80\% of the peak flux density. {\textbf{Left panel: }} The image is made following the method I. The dynamic range of the Stokes I image is $\sim$35, and that of the Stokes V image is $\sim$30. There are noise peaks at the 2\% level of the peak flux density in the Stokes I image. The residual Stokes V leakage is $\lesssim$5\%. \textbf{Right panel: }The image is made using P-AIRCARS. The dynamic range of the Stokes I image is $\sim$577, and that of the Stokes V image is $\sim$375, which is an order of magnitude better than that for method I. The first noise peak appears at the $0.8\%$ level of the peak flux density in the Stokes I image. We have detected another positive circularly polarized source (marked by the purple dashed box), which could not be detected in the image made using method I. This source has a circular polarization fraction $\sim3\%$. The residual Stokes V leakage is $\lesssim0.5\%$, which is also an order of magnitude improvement compared to that for method-I.}
    \label{fig:paircars_methodI}
\end{figure*}

Despite its scientific merits being well recognized, gyrosynchrotron emission from CMEs has been only been detected in a handful of cases until date, of which an even smaller fraction are at meter wavelengths \citep{bastian2001, Bain2014}.
Using AIRCARS, \cite{Mondal2020a} have recently demonstrated the ability to detect gyrosynchrotron emission from CME plasma and model spatially resolved spectra to estimate plasma parameters of the CME for a slow and unremarkable CME.
With further improved imaging from P-AIRCARS, it is now feasible to detect much fainter gyrosynchrotron emissions from CMEs and trace them out to higher coronal heights.
This capability will make it possible to use this powerful tool for routinely measuring the CME plasma parameters and magnetic fields in the middle and upper corona.

\subsection{Comparison with Method I}
Section \ref{sec:previous-attempts} discussed the earlier approach to  spectropolarimetric solar imaging using the MWA, which has been referred to as the method I in this text \citep{Patrick2019}, and its limitations. 
The robust first-principles-based polarization calibration approach of P-AIRCARS, on the other hand, ensures that all known instrumental effects are corrected for. The Stokes images delivered by P-AIRCARS are limited  primarily by the thermal noise of the data, and the rms noise seen in Stokes I and V images is significantly smaller than that seen in images from method I. 

A comparison between images delivered by the method I (left panel) and P-AIRCARS (right panel) is shown in Fig. \ref{fig:paircars_methodI}.
Both images have been made from same observation on 2014 October 24, 03:46:00 UTC at 159 MHz. 
The red contours represent the Stokes I emission. 
The method I image shows artifacts and noise peaks at $\sim$2\% of peak flux density. 
The largest noise peak in the P-AIRCARS image is at  $\sim$0.8\% of the peak flux density. 
The circular polarization percentage is shown by gray scales in regions where the Stokes I and V emissions are both detected with $> 3\sigma$ significance. 
The dynamic ranges of both the Stokes I and V images have increased by an order of magnitude -- for the Stokes I image from $\sim35$ to $\sim577$ and for the Stokes V image from $\sim30$ to $\sim375$. 

This improvement in dynamic range enables the detection of a much weaker Stokes V emission with $\sim$3\% circular polarization in the P-AIRCARS image (purple dotted box, right panel, Fig. \ref{fig:paircars_methodI}), which was not detected using method I, and the region over which the circular polarization is detected with confidence in the P-AIRCARS image is substantially larger than that in the method I image. $L_{residual}$ has been estimated to be $\sim5\%$ for the method I and $<|0.5|\%$ for P-AIRCARS. The peak circular polarization fraction of the image from method I is $\sim-22$\% and from P-AIRCARS $\sim-27\%$. Considering the $\sim5\%$ uncertainty on circular polarization fraction of method I \citep{Patrick2019}, both these values are consistent.
The very small $L_{residual}$ together with the high imaging dynamic range delivered by P-AIRCARS bodes very well for detection of the very low level of circularly polarized emission from the quiet-Sun thermal emission. 
This is also a very promising development for detection of polarized emission from gyrosynchrotron emission from the CME plasma.

\subsection{Polarization Images of the Sun Using P-AIRCARS}

\begin{figure*}[!t]
    \centering
    \includegraphics[trim={2cm 6.5cm 2.5cm 1.2cm},clip,scale=0.7]{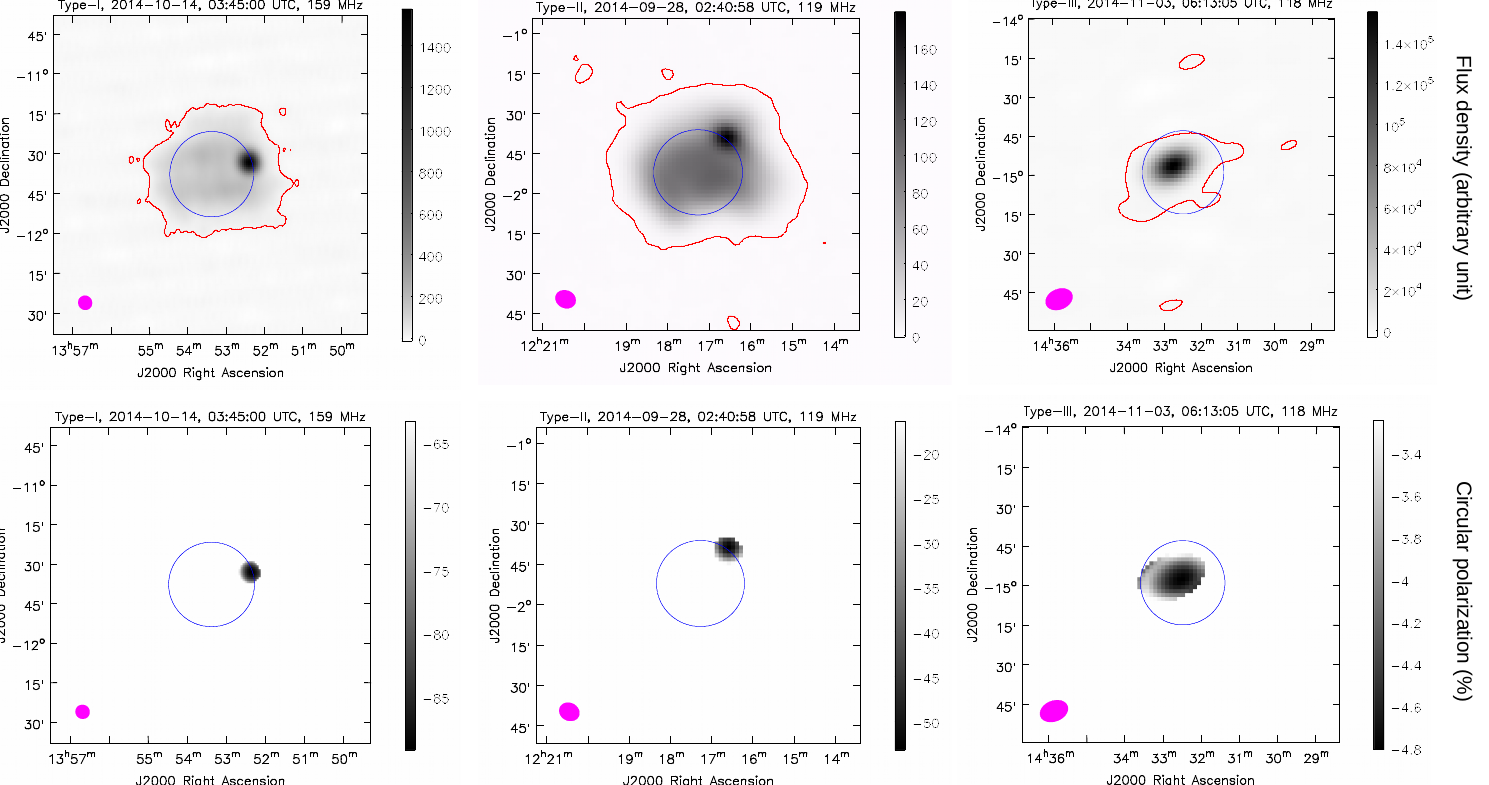}
    \caption{{\textbf{Example circular polarization images of types I,II, and III solar radio bursts.} Stokes I images are shown in the top panels and circular polarization fractions in the bottom panels.} Only the regions with $> 5\sigma$ detection in both Stokes I and Stokes V are shown in the bottom panels.
    The magenta ellipses at the bottom left of each panel show the PSF. The red contours in each image represent the first contour, which picks up the noise in Stokes I images. The optical disk of the Sun is shown by the blue circles. The residual Stokes V leakage in all of these images is $\lesssim1\%$. 
    {\textbf{Left panel: }} Circularly polarized emission from a type I noise storm at 159 MHz. The dynamic range of the Stokes I and Stokes V images is 788 and 518, respectively. The red contour is at 0.8\% of the peak Stokes I emission. The maximum circular polarization fraction is -89\%. {\textbf{Middle panel: }} Circular polarization from a type II solar radio burst at 119 MHz. The dynamic range of the Stokes I and Stokes V images is 900 and 950, respectively. The red contour is at 0.5\% of the peak Stokes I emission. The maximum circular polarization fraction is -53\%. {\textbf{Right panel: }} Stokes V image of a type III solar radio burst at 118 MHz. The dynamic range of the Stokes I and Stokes V images are 1200 and 233, respectively. The red contour is at the 0.8\% of the peak Stokes I emission. The maximum circular polarization fraction is -4.8\%.}
    \label{fig:radio_bursts}
\end{figure*}

Figure \ref{fig:radio_bursts} shows the Stokes I and V images for example type I, type II, and type III solar radio bursts. These images are made at the native time and frequency resolution, 0.5 s and 40 kHz, of the observation. We have chosen different contour levels for different images to show the first noise peak of that image. 
Circular polarization fractions are shown in gray scale over the regions where both Stokes I and Stokes V emissions are detected with $> 5\sigma$ significance. The peak circular polarization for these type I, type II and type III radio bursts are -89\%, -53\%, and -4.8\%, respectively. The Stokes V flux densities of these bursts are $\sim$ 70, 180, and 70 SFU, respectively.
In all cases we obtain images with DR varying between $\sim500$ and $1200$. 
The high DR of these images enables us to detect a significant part of the quiet-Sun emission in Stokes I, even the presence of the bright active emissions. 
Interestingly, the peak of the Stokes V emission from the type III burst is slightly but clearly displaced toward the southeast from the peak of Stokes I emission, while they are coincident for the type I and type II bursts shown.

These, along with the quiet-Sun data showcased in \ref{subsec:stress-test} span a large range along the flux density axis. The disk-integrated flux density of the quiet Sun is $\sim1.2\ \mathrm{SFU}$, and peak surface brightness is $0.2\ \mathrm{SFU}$ per beam at 80 $\mathrm{MHz}$. This demonstrates the capability of P-AIRCARS to produce high-DR images in a variety of different solar conditions.
Additionally, the residual instrumental polarization leakages of $\lesssim 1\%$ in all the observations shown in Fig. \ref{fig:radio_bursts}, represent an improvement approaching an order of magnitude over the $\sim5-10\%$ leakages obtained earlier \citep{Patrick2019,Rahman2020}.
To the best of our knowledge, these are the lowest to have been achieved for any meter-wavelength solar radio images and bring the exciting science target of measuring the weak circular polarization from the quiet Sun within reach.

\subsection{Heliospheric Measurements Using Background Radio Sources}
The Sun is the source with the highest flux density in the low-frequency sky, with even the quiet Sun spewing flux density exceeding $10^4-10^5\ \mathrm{Jy}$ in the MWA band.
Apart from a handful of sources whose flux density goes up to hundreds of $\mathrm{Jy}$, that of the bulk of other celestial sources lies in the range of a few $\mathrm{Jy}$ or weaker.
Imaging a reasonably dense grid of background sources in the presence of the Sun in the FoV, hence, imposes a large dynamic range requirement, which had not been met until recently.
The high dynamic-range of the images from P-AIRCARS now routinely allow us detect multiple background galactic and extragalactic radio sources even in the presence of the Sun.
An example is shown in Fig. \ref{fig:aircars_paircars_rms}, which is made over 2.56 $\mathrm{MHz}$ and 10 $\mathrm{s}$. The P-AIRCARS image shows a detection of 14 background sources with $\gtrsim5\sigma$ detection significance.
The closest of these is at $\sim$20 $\mathrm{R_{\odot}}$ from the Sun and has a flux density of 4.9 $\mathrm{Jy}$.
Another example is available in \citet{Kansabanik2022}, where the independently available flux densities of background sources were used for arriving at robust solar absolute flux density calibration for MWA solar observations. 
The rms noise in these images approaches that achieved by GLEAM, once
the excess system temperature due to the Sun is taken into account \citep{Kansabanik2022}.

Measurements of interplanetary scintillation (IPS) of background radio sources have long been used to measure the electron density, properties of turbulence and velocities of CMEs, and solar wind in the heliosphere \citep[e.g.][]{Coles1978,Manoharan1990, Jackson1998} and more recently for driving the boundary conditions for magnetohydroynamic models of the solar wind \citep{Yu2015}.
At the MWA, IPS observations have generally been done over small time windows while keeping the Sun at the null of the primary beam to avoid any contamination from solar emission \citep{morgan2018,Morgan2018a,Chettri2018}. 
By providing the ability to routinely make Stokes I images of background sources, P-AIRCARS brings us a step closer to removing this limitation and opens the possibility of performing IPS observations without necessarily requiring to have the Sun in a null of the MWA primary beam.

While IPS is a very useful remote-sensing technique and provides information complementary to what is available from other observations, it is not sensitive to heliospheric magnetic fields: the key driver of space weather phenomena.
By measuring the FR due to the heliospheric and/or CME plasma along the lines of sight to background linearly polarized sources or the diffuse galactic emission, radio observations provide the only known remote-sensing tool for measuring these magnetic fields.
This approach has been successfully implemented by using radio beacons from satellites \citep[e.g.][etc.]{bird1990, Jensen2013, Wexler2019} as background sources, and more interestingly also using astronomical sources \citep[e.g.][etc.]{Mancuso2000, kooi2017, Kooi2021}.
These observation were carried out at higher frequencies and using small FoV instruments, which can sample only a small part of the heliosphere at any given time.
Wide FoV instruments like the MWA can sample large swaths of the sky at any given time and can potentially track CMEs as they make way across the heliosphere.
Measurements of FR simultaneously for a large numbers of pierce points across the CME/heliosphere, open the very exciting possibility of constraining the models for CME/heliospheric magnetic fields using these data \citep{Bowman2013, Nakaraiakov2015}.
P-AIRCARS delivers precise polarization calibration and produces high-dynamic-range full-Stokes images and can already provide Stokes I maps of background sources.
We are pursuing the objective of demonstrating making similar full-Stokes maps of background sources and enabling these heliospheric FR measurements. 

\section{Conclusions and Future Work}\label{Conclusion} 
We have developed a robust and comprehensive state-of-the-art polarization calibration algorithm tailored for the needs of low-frequency solar observations.
P-AIRCARS builds on the learnings from our earlier Stokes I imaging pipeline \citep[AIRCARS;][]{Mondal2019} and uses the advantages endowed by the MWA design features to perform full polarimetric calibration without requiring dedicated observations of calibrator sources.
The key MWA design advantages in this context are the dense and compact core of the MWA array layout and its simple and well-characterized hardware.

Together these ensure that nearby antennas, the ones looking through essentially the same ionospheric patch, maintain a good degree of coherence.
A forthcoming paper will provide a detailed discussion and demonstration of these aspects (D. Kansabanik, 2022, in preparation).
All the P-AIRCARS images shown here were made without using any calibrator observations.

Polarization self-calibration was first demonstrated on simulated data by \citet{Hamaker2006}, but it had never been used for solar imaging.
This work presents the first demonstration of solar polarization self-calibration and its ability to achieve high-dynamic-range and high-fidelity full-Stokes images over a large range of solar conditions.
The residual Stokes leakages for these images are on par with the usual astronomical images. 

Though P-AIRCARS was developed with polarization calibration of the solar observations in mind, at its core the algorithm is general and does not impose any solar specific constraints. 
Its perturbative approach can be used for full Jones polarization self-calibration of the astronomical observations when a good initial sky model is available for a first-order calibration. 

The perturbative algorithm used in P-AIRCARS works well for homogeneous arrays like the MWA, where the ideal primary beam response of all antenna elements are essentially identical. 
MWA antenna elements are made of a total 16 bow-tie dipoles arranged in a 4$\times$4 grid \citep{Tingay2013}. 
The MWA beam is modeled assuming that all of the 16 dipoles are healthy \citep{Sokowlski2017}. 
It has been shown using satellite measurements that even when one or two of the  dipoles fail, it does not change the primary beam response close to its peak in a significant manner \citep{Line2018}. 
However, for precise polarization calibration being pursued here, these small changes do need to be accounted for.
Presently, in P-AIRCARS we reject all antenna elements with even a single dipole failure. 
Though it does lead to a loss of sensitivity, it is usually tolerable as the number of such elements is usually small.
However for science applications close to the edge of the sensitivity limits, e.g. detection of CME gyrosynchrotron emission, it can become important to retain the sensitivity offered by the elements with defective dipoles.
While the MWA beams can be modeled well for any subset of working dipoles \citep{Sokowlski2017}, it breaks the assumption of the array being a homogeneous one.
The implication for P-AIRCARS is that an image-based approach for corrections for primary beams is no longer tenable (Eq. \ref{eq:beamcor_2}).
One must then use a class of algorithms referred to in the literature as {\it projection} algorithms, which can correct for image plane effects in the visibility domain. 
These algorithms can be used for correcting artifacts arising from a wide range of causes, ranging from the so called {\it w term} to the antenna-to-antenna differences in primary beams even for an array with identical elements and ionospheric phase screens.
The algorithm of relevance is the one referred to as the {\it aw projection} algorithm \citep{Jagannathan2017,Jagannathan2018,Sekhar2021}.
It applies baseline-based corrections for primary beams in the visibility domain and is computationally very intensive. 
As efficient implementations of such algorithms become available and  computational capacity available to us grows, it will become interesting to explore their use for scientifically interesting datasets with significant numbers of dipole failures to squeeze the most out of these data.

P-AIRCARS has been developed with the future SKA in mind. It can easily be adapted for unsupervised generation of high-fidelity high-dynamic-range full-Stokes images for the SKA and other similar instruments with a dense central core. 
Our intent is to make a stable and mature implementation of P-AIRCARS available to the larger solar physics community.
Producing high-quality solar radio interferometric images involves a steep learning curve, and its practise has remained limited to a small subset of the solar physics community.
It is our belief that the lack of availability of a robust tool suitable the nonspecialist has long limited the use of radio observations in solar studies.
We envisage that P-AIRCARS will prove to be a very useful tool for the solar and heliospheric physics community in times to come by filling this gap and making high quality full Stokes solar radio imaging accessible.

\facilities{Murchison Widefield Array \citep[MWA][]{lonsdale2009,Tingay2013,Wayth2018}}

\software{astropy \citep{price2018astropy}, matplotlib \citep{Hunter:2007}, Numpy \citep{Harris2020}, SciPy \citep{Scipy2020}, CASA \citep{mcmullin2007}, CubiCal \citep{cubical2018,Cubical_robust2019}}

\begin{acknowledgments}
This scientific work makes use of the Murchison Radio Astronomy Observatory (MRO), operated by the Commonwealth Scientific and Industrial Research Organisation (CSIRO). We acknowledge the Wajarri Yamatji people as the traditional owners of the Observatory site. Support for the operation of the MWA is provided by the Australian Government's National Collaborative Research Infrastructure Strategy (NCRIS), under a contract with Curtin University administered by Astronomy Australia Limited. We acknowledge the Pawsey Supercomputing Centre, which is supported by the Western Australian and Australian Governments. D.K. gratefully acknowledge Barnali Das (University of Delaware, Newark, USA) for useful discussions, suggestions and also for providing a beautiful name of the pipeline. D.K. gratefully acknowledge John Morgan (Curtin University, Australia) for providing useful information about software packages used previously for MWA polarization calibration at the early stage of this project. D.K. acknowledges the discussion with Xiang Zhang (ICRAR, Australia) for useful discussion about the cross-hand phase.  We also gratefully acknowledge the helpful comments from the anonymous referee, which have helped to improve the clarity and the presentation of this work. D.K. and D.O. acknowledge the support of the Department of Atomic Energy, Government of India, under the project No. 12-R\&D-TFR-5.02-0700. S.M. acknowledges partial support by USA NSF grant AGS-1654382 to the New Jersey Institute of Technology.
\end{acknowledgments}

\bibliography{algorithm.bib}{}
\bibliographystyle{aasjournal}

\end{document}